\documentclass[journal,A4]{IEEEtran}

\usepackage{cite}
\usepackage{graphicx}
\usepackage{amssymb,amsmath} 
\usepackage{multirow}
\usepackage{booktabs}
\usepackage{url}
\makeatletter
\let\MYcaption\@makecaption
\makeatother

\usepackage[font=footnotesize]{subcaption}

\makeatletter
\let\@makecaption\MYcaption
\makeatother

\begin{document}

%\title{Validation of physics methods for Monte Carlo simulation of the photoelectric effect}
\title{Evolutions in photoelectric cross section calculations and their validation}

\author {Tullio Basaglia, Maria Grazia Pia and Paolo Saracco% <-this % stops a space
\thanks{Manuscript received 22 November 2019.}% <-this % stops a space
%\thanks{This work has been partly funded by ?}% <-this % stops a space
\thanks{T. Basaglia is with CERN Scientific Information Service, CH-1211, Geneva, Switzerland (e-mail: Tullio.Basaglia@cern.ch).}
\thanks{M. G. Pia and P. Saracco are  with INFN Sezione di Genova, Via Dodecaneso 33, I-16146 Genova, Italy 
	(phone: +39 010 3536328,  e-mail: 
	MariaGrazia.Pia@ge.infn.it, Paolo.Saracco@ge.infn.it).}
}

\markboth{IEEE Transactions on Nuclear Science,~Vol.~x, No.~n, Month~Year}%
{Evolutions in photoelectric cross section calculations and their validation}

\maketitle

\begin{abstract}
This paper updates and complements a previously published evaluation of
computational methods for total and partial cross sections, relevant to modeling
the photoelectric effect in Monte Carlo particle transport.
It examines calculation methods that have become available since the publication
of the previous paper, some of which claim improvements over previous
calculations; it tests them with statistical methods against the same sample of experimental data
collected for the previous evaluation.
No statistically significant improvements are observed with respect to
the calculation method identified in the previous paper as the state of the art
for the intended purpose, encoded in the EPDL97 data library.
Some of the more recent computational methods exhibit significantly lower capability 
to reproduce experimental measurements than the existing alternatives.
\end{abstract}
\begin{keywords}
Cross sections, Photoionization, Monte Carlo simulation
\end{keywords}

% -----------------------------------------------------------------------------------------

\section{Introduction}
\label{sec_intro}

\PARstart{P}{hotoionization} has attracted theoretical and experimental interest
for many decades, which is reflected in the continuing expansion of theoretical
and empirical methods to calculate the cross sections associated with this process.

A previous publication \cite{tns_photoel} examined several total
and partial photoionization cross section calculations with respect to a wide
collection of experimental measurements to assess the state of the art for
modeling the photoelectric effect in Monte Carlo particle transport.
The validation tests documented in \cite{tns_photoel} identified the tabulations
in the Evaluated Photon Data Library (EPDL), 1997 version \cite{epdl97}, based
on Scofield's 1973 calculations \cite{scofield_1973}, as the
calculation method best reproducing experimental cross sections.
An overview of EPDL and Scofield's calculations is available in \cite{tns_photoel};
a recent review can be found in \cite{andreobook}.

Other calculation methods have become available since completion of the
tests documented in \cite{tns_photoel}, either based on theoretical approaches
or on empirical parameterizations: calculations by Sabbatucci and Salvat
\cite{sabbatucci_2016}, modified tabulations in Penelope 2014
\cite{penelope2014}, a modified version of EPDL \cite{epdl2017} released in 2018,
% within the EPICS2017 \cite{epics2017} data collection, 
new parameterizations \`a la
Biggs-Lighthill \cite{biggs1, biggs2, biggs3}, developed in the context of
GeantV \cite{geantv} and adopted in Geant4 \cite{g4nim, g4tns, g4nim2}.

This paper updates the computational scenario documented in the previous one by
evaluating the newly available modeling options with respect to the same
collection of experimental data retrieved from the literature.
The cross section calculations subject to test are summarized in Table
\ref{tab:models}.

% that  were used for the validation tests documented in \cite{tns_photoel}.

%This paper reports an update to the conclusions of the 
%previous publication by assessing the state of the art of photoelectric cross 
%sections for particle transport among the wider set of calculation methods 
%available to date.

% -----------------------------------------------------------------------------------------

\section{Evolutions in photoelectric cross sections}
\label{sec:newmethods}

Like \cite{tns_photoel}, this paper focuses on
calculations relevant to the simulation of the photoelectric effect in general-purpose 
Monte Carlo codes, which typically deal with single ionisation of neutral atoms.
%As in \cite{tns_photoel}, 
The analysis concerns cross sections that are available as tabulations of precalculated
values or can be expressed as simple analytical formulations, suitable for
implementation in particle transport codes.

An overview of calculation methods of photoelectric cross sections 
is summarized in \cite{tns_photoel}.
A few relevant developments that became available at a
later stage are briefly discussed below.

It is worthwhile to stress that all the computational methods considered in
this paper assume free atoms.
This assumption is also adopted by all major general-purpose particle transport
codes; it is liable to introduce inaccuracies for molecules and solids,
especially near absorption edges, due to aggregation effects on the atomic
potential and EXAFS (extended X-ray absorption fine structure).
Therefore, some discrepancies observed with respect to experimental measurements
are to be ascribed to the common underlying theoretical approach, rather than to
specific deficiencies of any of the computational methods tested.

% -----------------------------------------------------------------------------------------
\subsection{Theoretical calculations}

Sabbatucci and Salvat \cite{sabbatucci_2016} 
%reformulated the theory of the photoelectric effect; they 
applied 
first-order perturbation theory and Dirac one-electron wave functions for the
Dirac-Hartree-Fock-Slater (DHFS) self-consistent potential of  free
atoms,
%the Dirac-Hartree-Fock-Slater (DHFS) calculation method, 
which are also
the basis of EPDL97 tabulations, to generate a database of atomic cross
sections for the elements of the periodic table, including total cross sections
and partial cross sections for the K shell and for L, M and N subshells with
binding energies greater than approximately 50 eV.

The  calculations are implemented in a code named \textit{PHOTACS}.
Additional effects are accounted for by a post-processing code,
identified as \textit{PHOTACS-PP}: the contribution from excitation to bound
levels, the effect of the atomic level widths and a normalization correction.
The possible computational configurations are listed in Table \ref{tab:sabid}.

The normalization correction is based on a method devised by Pratt
\cite{pratt_1960b, pratt_1973, pratt_1973_err}; it aims to address possible
inaccuracies of the adopted central potential by applying an energy-independent
correction factor, derived from a more elaborate atomic model, to the subshell
cross sections calculated from the DHFS potential.
This correction exploits calculations according to the multi-configuration
Dirac–Fock \textit{MCDF} code by Desclaux \cite{desclaux_1975, desclaux_1977}.
%Calculations based on approximate independent-electron models, such as the DHFS model, are affected by  . A simple strategy to account for inaccuracies in the atomic potential is provided by the normalization screening approximation of Pratt and co-workers (Pratt, 1960a, Schmickley and Pratt, 1967, Pratt and Tseng, 1972, Pratt et al., 1973). According to this approximation, the subshell cross sections calculated from the DHFS potential and from a more elaborate atomic model (e.g.., the multi-configuration Dirac–Fock self-consistent model implemented in the program of Desclaux, 1975, Desclaux, 1977) differ essentially by a constant factor, which is equal to the ratio of electron densities near the nucleus. That is, in principle, one can improve the DHFS cross sections by multiplying them by an energy-independent factor, which is readily obtained from atomic-structure calculations that are more elaborate than the DHFS ones.

%Atomic states are described within the independent-electron approximation, with
%bound and free one-electron orbitals that are solutions of the Dirac equation
%with the Dirac–Hartree–Fock–Slater self-consistent potential of the ground-state
%configuration. Detailed derivations are given of subshell cross sections for
%both excitation to discrete bound levels and ionization.
%
%(Section 2.3), (Section 2.5), and t(Section 2.6) are accounted for by a
%post-processing program, named photacs-pp, which reads the tables of numerical
%cross sections generated by photacs

The database derived from \cite{sabbatucci_2016} is used in the Penelope
Monte Carlo code \cite{sabbatucci_2016, salvat_2018}.

% -----------------------------------------------------------------------------------------
\subsection{Penelope 2014 cross sections}

The latest version of Penelope publicly available at the time of writing this paper is
Penelope 2014 \cite{penelope2014}, which was released in 2015.

According to the associated documentation \cite{penelope2014}, the photoelectric
cross sections tabulated in Penelope 2014 derive from calculations performed by
a computer code named \textit{PHOTOABS}, which is based on the same
Dirac-Hartree-Slater-Fock method as Scofield's 1973 calculations, but
``implements more accurate numerical algorithms'' and computes them over a
denser grid of energies.

Penelope 2014 documentation \cite{penelope2014} cites an associated reference
to be published in 2014 as a source of further details.
No pertinent paper appears to be published in 2014; \cite{sabbatucci_2016},
which was published in 2016, states that the database generated by the cross
section calculations it describes was adopted in Penelope 2014.

The cross sections associated with Penelope 2014 were calculated for photon
energies from the ionization threshold up to 1 GeV using the DHFS approach;
% by the \textit{PHOTOABS} code
they were extrapolated to energies higher than the calculation cut-off by means
of Pratt's extrapolation formula \cite{pratt_1960a}, shifted in energy to have
the absorption edges coinciding with the experimental subshell ionization
energies given by Carlson \cite{carlson_1975} and renormalized using MCDF/DHFS
density ratios.
The  normalisation correction is meant ``to correct some inaccuracies
of the DHFS wave functions''.

%An updated version of Penelope, illustrated in training courses but not yet
%available from NEA Data Bank, uses updated cross sections calculated by
%\textit{PHOTACS}.

The tabulations of photoelectric cross section distributed with previous
versions of Penelope (up to Penelope 2011 \cite{penelope2011}) were based on EPDL97.

% -----------------------------------------------------------------------------------------
\subsection{EPICS2017 cross sections}

A new version of EPDL was released in 2018 within ENDF/B-VIII.0 \cite{endfb8} 
and in a collection of data libraries called EPICS2017 \cite{epics2017}.
It accounts for a modification
of the binding energies previously encoded in EADL \cite{eadl, eadl2017};
according to \cite{epdl2017}, the photoelectric cross sections in EPICS2017 were
obtained by modifying the EPDL97 \cite{epdl97} tabulations by smoothly
interpolating or extrapolating the values close to absorption edges consistently
with the modified binding energies.

In addition, photoelectric cross sections are tabulated in EPICS2017 according
to a finer grid of energies than in EPDL97 to support linear interpolation of
the data, while logarithmic interpolation was recommended for EPDL97.

Deficiencies in version control were observed \cite{tns_epics2017} in the
distribution of EPICS2017: different content has been distributed in various
instances under the same identifier of EPICS2017, therefore it is impossible to
identify EPICS2017 photoelectric cross sections univocally.
The cross sections examined in this paper correspond to those available in ENDF
format within ENDF/B-VIII.0 \cite{endfb8}; no further versions of these cross
sections are mentioned in the ENDF/B-VIII.0 errata \cite{endfb8_errata}.
Different versions of photoelectric cross sections were released in ENDL format 
\cite{epics2017} in January and February 2018, both identically identified as
EPICS2017; the latter appears equivalent to the tabulations included in
ENDF/B-VIII.0.

% -----------------------------------------------------------------------------------------
\subsection{Parameterized cross sections}
\label{sec:param6}

A new parameterization of photoelectric cross sections is available in Geant4 version 10.5.  
% and is described in the associated Geant4 documentation.
It is inspired by Biggs-Lighthill \cite{biggs1, biggs2, biggs3} empirical expression of total
photoelectric cross sections
\begin{equation}
\sigma_{ij} = \frac{A_{ij1}}{E} + \frac{A_{ij2}}{E^2} + \frac{A_{ij3}}{E^3} + \frac{A_{ij4}}{E^4} 
\label{eq_biggs}
\end{equation}
for element $i$ and energy range $j$.
The modified parameterization, identified as ``Param6'' in Table {tab:models}, is formulated as
\begin{equation}
\sigma_{ij} = \frac{A_{ij1}}{E} + \frac{A_{ij2}}{E^2} + \frac{A_{ij3}}{E^3} + \frac{A_{ij4}}{E^4} + \frac{A_{ij5}}{E^5} + \frac{A_{ij6}}{E^6} 
\label{eq_biggs6}
\end{equation}
over two energy ranges (``low'' and ``high''), whose limits depend on the atomic
number.
According to the Geant4 10.5 documentation  \cite{g4physrefmanual10.5}, the
coefficients appearing in equation (\ref{eq_biggs6}) are calculated by fitting
EPICS2014 \cite{epics2014} tabulations, which are identical to EPDL97 ones \cite{tns_epics2017}.
Although the bibliographical reference to the fitted tabulations in \cite{g4physrefmanual10.5} refers to EPICS2017 at
the time of writing this paper,  it specifies that it was accessed on 26 October 2017,
while EPICS2017 was released in January 2018; presumably, the online content of the reference was modified 
after the release of Geant4 10.5 documentation.

The low energy end of applicability of the parameterization is approximately
5~keV; below this limit, which varies according to the atomic number, the cross
section is computed by interpolation of EPDL tabulations.
Equation  (\ref{eq_biggs6}) can calculate total and shell cross sections.

This parameterized model is used for the simulation of the photoelectric effect by
the Geant4 ``low energy Livermore'' model; it replaces a previous implementation, which
calculated total and partial photoelectric cross sections by interpolating EPDL97 
tabulations.
%
%G4Param6
%
%To the best of our efforts, we could not identify a proper reference in the literature
%documenting the parameterized photoelectric cross section model
%implemented in Geant4.
%
%For convenience, this paper also reports the results based on the original
%Biggs-Lighthill parameterization, which were published in \cite{tns_photoel}.

%\subsection{Tabulations}
%
%EPDL, EPICS 2017, Penelope
%
%For convenience, this paper reports the results of both EPDL versions: the
%original EPDL97 and the version contained in EPICS 2017.
% -----------------------------------------------------------------------------------------

% -----------------------------------------------------------------------------------------
\begin{table*}[htpb]
  \centering
  \caption{Calculation methods of photoelectric cross sections considered in this study}
    \begin{tabular}{llcrrrc}
   \toprule
    \textbf{Method}  &{\bf Identifier} & \multicolumn{2}{c}{\textbf{Energy range}} & \multicolumn{2}{c}{\textbf{Z range}} & \textbf{Shell} \\
    \midrule
   EPDL, 1997 version			\cite{epdl97}			& EPDL97		&  10 eV 	& 100 GeV  	&  1	& 100 	& all \\
   EPDL as in EPICS2017		\cite{epdl2017}			& EPICS2017	&  10 eV   	& 100 GeV 	&  1	& 100	& all \\
  Biggs-Lighthill parameterization \cite{biggs3} 			& Biggs		&  10 eV 	& 100 GeV 	&  1	& 100	& \\
  Geant4 10.5 parameterization						& Param6		&  $\sim$5 keV 	& 100 TeV &  1	& 100	& all \\
  Penelope 2011				\cite{penelope2011}		& Pen2011		&  50 eV 	& 1 GeV 		&  1	& 99 		& K, L, M, N \\
  Penelope 2014				\cite{penelope2014}		& Pen2014		&  50 eV 	& 1 GeV 		&  1	& 99 		& K, L, M, N \\
  Sabbatucci-Salvat 			\cite{sabbatucci_2016}	& Sab[\textit{option}]		& Ionization threshold		& 10 GeV			&  1	& 99		& all \\
    \bottomrule
    \end{tabular}%
  \label{tab:models}%
\end{table*}%

% -----------------------------------------------------------------------------------------
\section{Validation method}

The validation \cite{ieee_vv} process applies the methodology described in \cite{tns_photoel}:
comparison with experimental data to determine the compatibility of each
calculation model with experiment, and categorical data analysis to identify the
state of the art among the calculations subject to test.

The following subsections illustrate the additional computational methods
evaluated in this paper and briefly summarize the main features of the
validation method.
The reader is referred to \cite{tns_photoel} for further details.

%% -----------------------------------------------------------------------------------------
%\begin{table*}[htpb]
%  \centering
%  \caption{Calculation methods of photoelectric cross sections considered in this study}
%    \begin{tabular}{llcrrrc}
%   \toprule
%    \textbf{Method}  &{\bf Identifier} & \multicolumn{2}{c}{\textbf{Energy range}} & \multicolumn{2}{c}{\textbf{Z range}} & \textbf{Shell} \\
%    \midrule
%   EPDL97	\cite{epdl97}				& EPDL97		&  10 eV 	& 100 GeV  	&  1	& 100 	& all \\
%   EPICS2017	\cite{epdl2017}			& EPICS2017	&  10 eV   	& 100 GeV 	&  1	& 100	& all \\
%  Biggs-Lighthill parameterization \cite{biggs3} & Biggs	&  10 eV 	& 100 GeV 	&  1	& 100	& \\
%  Geant4 10.5 parameterization			& Param6		&  $\sim$5 keV 	& 100 TeV &  1	& 100	& all \\
%  Penelope 2011	\cite{penelope2011}		& Pen2011		&  50 eV 	& 1 GeV 		&  1	& 99 		& K, L, M, N \\
%  Penelope 2014	\cite{penelope2014}		& Pen2014		&  50 eV 	& 1 GeV 		&  1	& 99 		& K, L, M, N \\
%  Sabbatucci-Salvat \cite{sabbatucci_2016}	& Sab[\textit{option}]		& Ionization threshold		& 10 GeV			&  1	& 99		& K, L, M, N \\
%    \bottomrule
%    \end{tabular}%
%  \label{tab:models}%
%\end{table*}%

% -----------------------------------------------------------------------------------------
\subsection{Cross section calculations}

%The cross section calculations subject to test are summarized in Table
%\ref{tab:models}.
The cross section calculations examined in this paper include some relevant computational methods previously evaluated in
\cite{tns_photoel}, along with the recent calculations mentioned in Section
\ref{sec:newmethods}, to facilitate the comparison of the evolution of similar
computational approaches.

A few configuration options can be applied to the cross sections calculated by
Sabbatucci and Salvat \cite{sabbatucci_2016} through different settings of the
\textit{PHOTACS-PP} code; they are listed in Table \ref{tab:sabid}  with
the corresponding identifiers used in in this paper.
They consist of incorporating the excitation to bound levels in the cross
section calculation along with ionisation, accounting for finite level width, and
applying Pratt's renormalization correction; additionally, the
\textit{PHOTACS-PP} code offers the choice of two compilations of atomic binding
energies of empirical origin, by Carlson \cite{carlson_1975} and Williams
\cite{williams_2011}, respectively, as an alternative to the default ones.
If atomic binding energies other than the default ones are chosen, the PHOTACS-PP code
adjusts the energy corresponding to the cross section calculated with the default value to let
the absorption edge coincide with the selected ionization energy.
Details about the PHOTACS-PP code are documented in \cite{sabbatucci_2016}.

\begin{table}[htbp]
  \centering
  \caption{Options for the calculation of Sabbatucci-Salvat photoelectric cross sections }
    \begin{tabular}{lcccc}
\toprule
\multirow{2}[0]{*}{ {\bf Identifier} }	& {\bf Binding} 	& {\bf Pratt} 		& {\bf Level width} 	&  \multirow{2}[0]{*}{ {\bf Excitation} } \\
            						& {\bf energy} 	& {\bf correction} 	& {\bf correction} 	 & \\
\midrule
    SabCar 		& \multirow{4}[0]{*}{Carlson} 	& no	& no 	& no \\
    SabWCar 	&       					& no  & yes & no \\
    SabPCar 	&       & yes   & no & no \\
    SabWPCar 	&       & yes   & yes & no \\
\midrule
    SabWil 		& \multirow{4}[0]{*}{Williams} & no    & no & no \\
    SabWWil		&       & no    & yes 	& no \\
    SabPWil 		&       & yes   & no 	& no \\
    SabWPWil 	&       & yes   & yes 	& no \\
\midrule
    SabDef 		& \multirow{4}[0]{*}{Default} & no    & no & no \\
    SabWDef 	&       & no    & yes 	& no \\
    SabPDef 	&       & yes   & no 	& no \\
    SabWPDef 	&       & yes   & yes 	& no \\
\midrule
    SabECar 	& \multirow{4}[0]{*}{Carlson} & no    & no & yes \\
    SabEWCar 	&       & no    & yes  	& yes \\
    SabEPCar 	&       & yes   & no  	& yes \\
    SabEWPCar 	&       & yes   & yes  	& yes \\
\midrule
    SabEWil 		& \multirow{4}[0]{*}{Williams} & no    & no  & yes \\
    SabEWWil 	&   	& no    	& yes  	& yes \\
    SabEPWil 	&    	& yes   	& no		& yes \\
    SabEWPWil 	&	& yes   	& yes  	& yes \\
\midrule
    SabEDef 	& \multirow{4}[0]{*}{Default} & no    & no  & yes \\
    SabEWDef 	&       & no    & yes	& yes \\
    SabEPDef 	&       & yes   & no  	& yes \\
    SabEWPDef 	&       & yes   & yes  	& yes \\
\bottomrule
    \end{tabular}%
  \label{tab:sabid}%
\end{table}%

The parameterized cross section model described in Section \ref{sec:param6} was
refactored \cite{refactoring} consistently with the software design documented
in \cite{tns_photoel} to facilitate the validation process.
Tabulated cross sections were interpolated logarithmically or linearly, according to
the recommendations of use of the respective data libraries.

% -----------------------------------------------------------------------------------------
\subsection{Experimental data}

The experimental data samples used for the validation tests reported in this
paper are the same as in \cite{tns_photoel}.
Details can be found in Tables II and III  of \cite{tns_photoel}.
%The corresponding references can be found in the bibliography of \cite{tns_photoel}.

Test cases were defined exactly as in \cite{tns_photoel} for the comparison of
total cross sections.

Test cases for the comparison of K shell data were defined for this analysis by
grouping the experimental data mainly at fixed energies, i.e. assembling data
samples consisting of the measurements performed by an experiment at a given
energy for various target elements, to increase the power of the tests by means
of larger sample sizes, while in the previous paper the test cases were mainly
defined on the basis of the atomic number.
However, it is worthwhile to stress that the experimental data set is the
same as in the analysis documented in \cite{tns_photoel} and that the validation
results obtained with either set of test cases are statistically consistent.
% with those reported in \cite{tns_photoel}.
% (they are based on the same experimental data indeed); 
%the different grouping contributes to enhance the power of the test comparing
%experimental and calculated cross sections, which are performed over larger data
%samples, while the reduced number of K shell test cases is reflected in the categorical 
%data analysis
%allows a more effective
%evaluation of scenarios involving absorption edges, which are of special
%interest in this paper, since some recent cross section calculations claim
%improved accuracy close to absorption edges.

% -----------------------------------------------------------------------------------------
\subsection{Data analysis method}
\label{sec:anal}

The data analysis method is extensively documented in \cite{tns_photoel}; it is
briefly summarized here to facilitate the comprehension of the results 
reported in Section \ref{sec:results}.

The validation process encompasses two stages: the first tests the compatibility
between the cross sections calculated by each simulation model and experimental
data, while the second determines whether the models exhibit any significant
difference in compatibility with experiment.

Both analyses employ statistical methods: the first stage applies the $\chi^2$
goodness-of-fit test \cite{bock} to appraise the compatibility of calculated and
experimental cross sections; the second applies a set of tests to
contingency tables based on the outcome of the first stage. 
Exact tests (Fisher \cite{fisher}, Boschloo \cite{boschloo}, Z-pooled
\cite{suissa} and Barnard \cite{barnard} in the CSM approximation
\cite{barnard_csm}) are used in the analysis of contingency tables; Pearson's
$\chi^2$ test \cite{pearson} is also used, when the number of entries in the
cells justifies its applicability (i.e., it is greater than 5 \cite{lyonsbook}).

The level of significance applied in all the tests to reject the null hypothesis is 0.01,
unless otherwise specified.

The analysis has been designed to ensure that the tests have adequate power to
observe sizeable effects of physical interest, given the available sample of experimental data.
The need to retain adequate power hinders more detailed analyses, such as a
thorough assessment of the incompatibility of the calculation methods with
experiment as a function of the atomic number and of energy.

The robustness of the results reported in the following sections has been
investigated with respect to several factors, e.g. with respect to measurements
with different precision and to different formulations of the test statistic in
both stages of the analysis; a few relevant details are discussed below.
%(for instance, the hint of energy dependence of the
%different incompatibility with experiment of EPDL97 and Sabbatucci-Salvat
%calculations involving Pratt's correction).

The analysis reported in this paper used the R software system \cite{R}, version 3.6.1.
%It is worthwhile to stress that  these only have a qualitative role.

% -----------------------------------------------------------------------------------------
\section{Results}
\label{sec:results}

The main results of the validation process are reported in the following subsections
for total and partial cross sections.

% near-edge absorption structures produced by molecular or crystalline ordering (e.g., extended x-ray absorption fine-structure) are ignored

% -----------------------------------------------------------------------------------------
\subsection{Total cross sections}
\label{sec_resulttot}

\begin{figure}
\centerline{\includegraphics[angle=0,width=8.0cm]{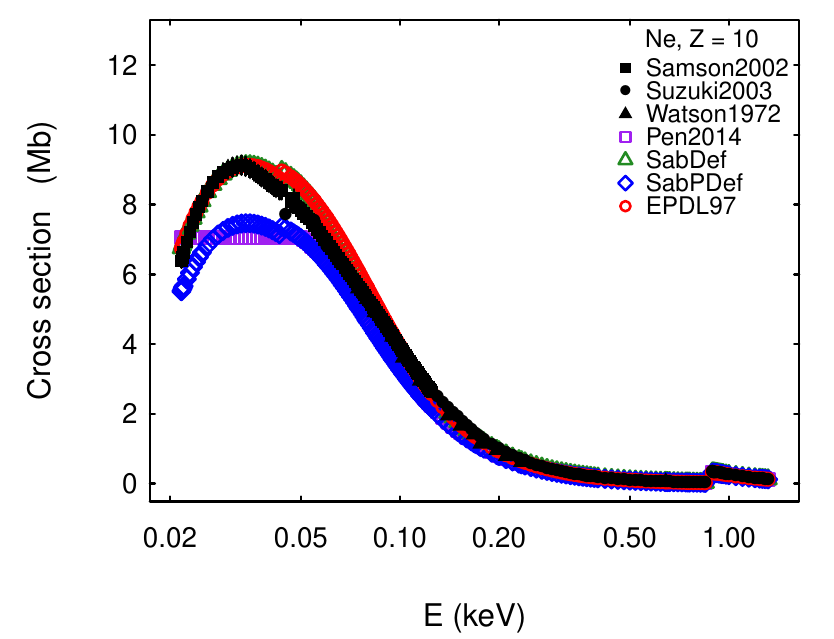}}
\caption{Total photoelectric cross section for neon as a function of photon energy.
The experimental references are listed in the bibliography of \cite{tns_photoel}.}
\label{fig_tot10}
\end{figure}

\begin{figure}
\centerline{\includegraphics[angle=0,width=8.0cm]{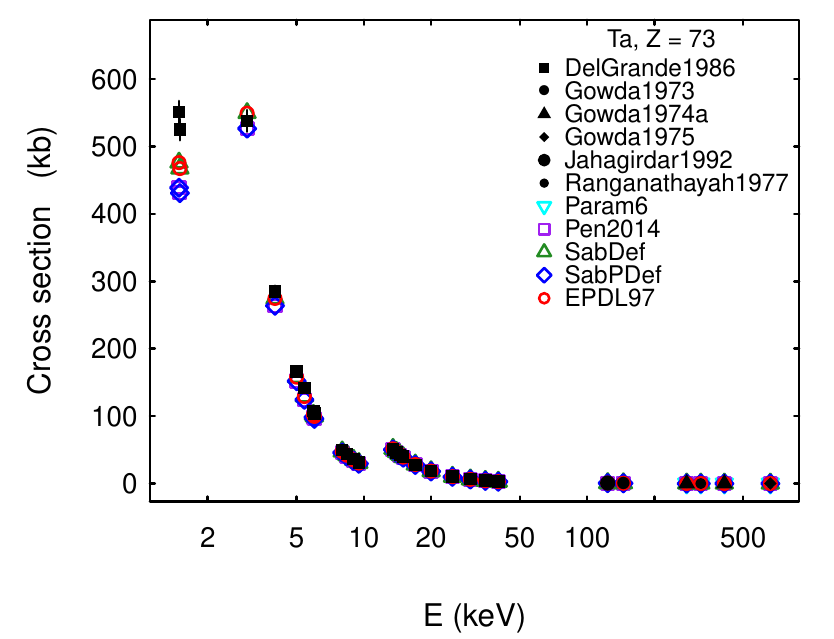}}
\caption{Total photoelectric cross section for tantalum as a function of photon energy.
The experimental references arelisted  in the bibliography of \cite{tns_photoel}.}
\label{fig_tot73}
\end{figure}

\begin{figure}
\centerline{\includegraphics[angle=0,width=8.0cm]{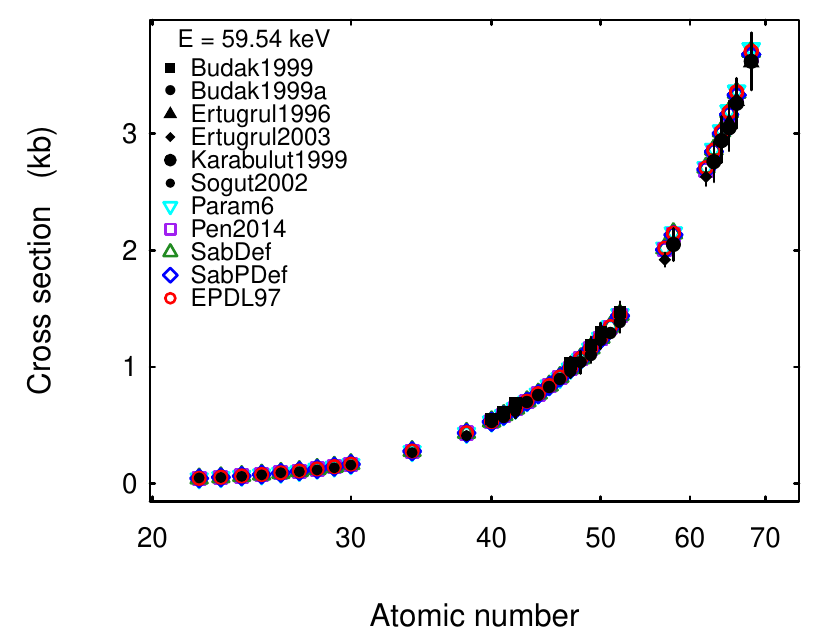}}
\caption{Total photoelectric cross section at 59.54~keV as a function of the atomic number.
The experimental references are in the bibliography of \cite{tns_photoel}.}
\label{fig_tote59}
\end{figure}

\begin{figure}
\centerline{\includegraphics[angle=0,width=8cm]{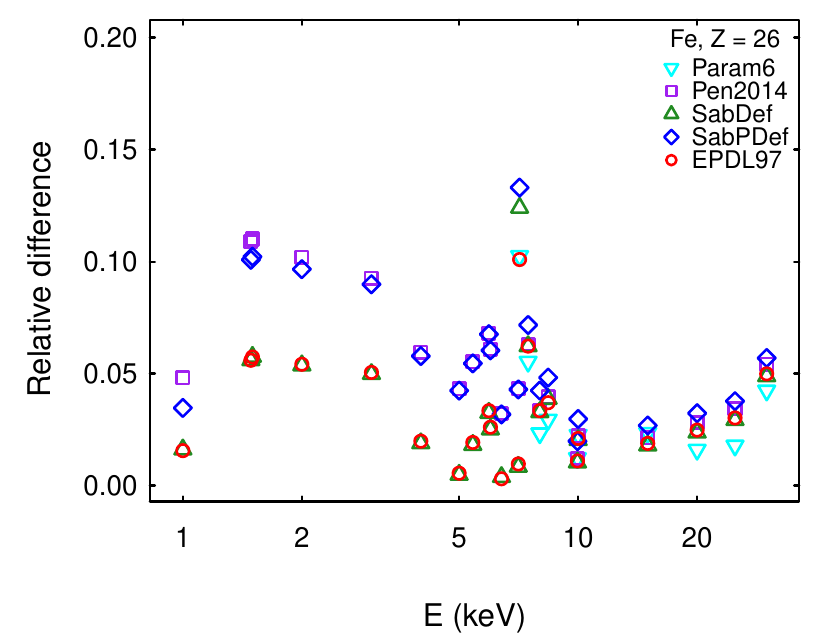}}
\caption{Absolute relative difference between calculated and experimental total photoelectric cross section for iron as a function of photon energy.
The experimental references  are documented in Table II of \cite{tns_photoel}.
Error bars are omitted to facilitate the appraisal of relevant features of the data points.}
\label{fig_totdiff26}
\end{figure}

\begin{figure}
\centerline{\includegraphics[angle=0,width=8cm]{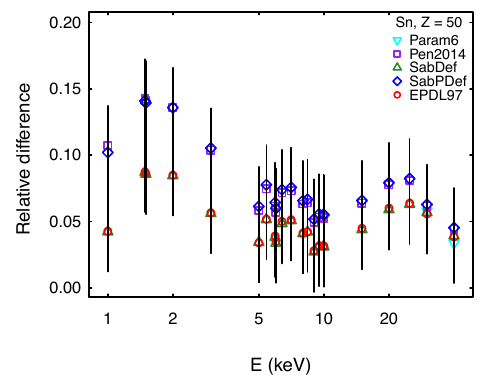}}
\caption{Absolute relative difference between calculated and experimental total photoelectric cross section for tin as a function of photon energy.
The experimental references  are documented in Table II of \cite{tns_photoel}.
%Error bars are omitted to facilitate the appraisal of relevant features of the data points.
}
\label{fig_totdiff50}
\end{figure}

\begin{figure}
\centerline{\includegraphics[angle=0,width=8cm]{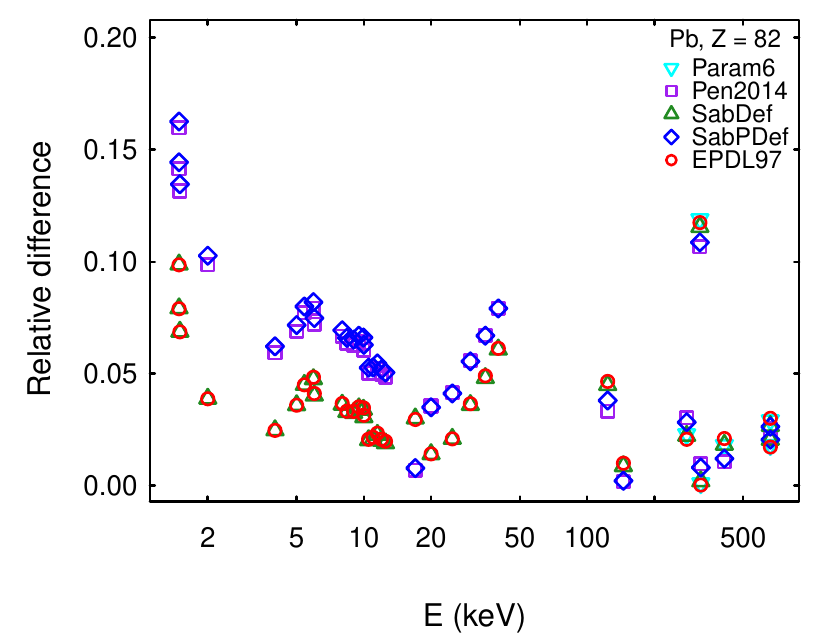}}
\caption{Absolute relative difference between calculated and experimental total photoelectric cross section for lead as a function of photon energy.
The experimental references  are documented in Table II of \cite{tns_photoel}.
Error bars are omitted to facilitate the appraisal of relevant features of the data points.}
\label{fig_totdiff82}
\end{figure}

Figs. \ref{fig_tot10}-\ref{fig_totdiff82} represent  qualitative examples
of how EPDL97 and some more recent computational methods reproduce experimental
total photoelectric cross sections.
Error bars are omitted in Figs. \ref{fig_totdiff26} and \ref{fig_totdiff82} to facilitate the 
appraisal of relevant features in the plots, i.e. the visible  discrepancies of 
data points involving Pratt's correction and the similarities concerning other calculations.
One can qualitatively distinguish some differences in the data plotted in
Figs.\ref {fig_tot10}-\ref{fig_totdiff82}: they mainly concern Sabbatucci-Salvat
cross sections including the correction à la Pratt, while other data sets appear
hardly distinguishable.

The investigation of the compatibility of calculated and measured total cross
sections focuses on energies greater than 100~eV and above approximately 5~keV.
The former energy range reflects the recommendation of applicability of EPICS2017 
photon cross sections \cite{epdl2017}; the latter corresponds
to the domain of application of the parameterized model described in section
\ref{sec:param6} adopted in Geant4 low energy electromagnetic package.

The outcome of the $\chi^2$ test above 100~eV is summarized in Table
\ref{tab:efftot100}, where ``fail'' and ``pass'' identify the number of test cases
where the hypothesis of compatibility between calculated and experimental data
distributions is rejected or not rejected, respectively. 
For convenience, the table reports a variable named ``efficiency'', which
represents the fraction of test cases where the null hypothesis is not rejected.
Similarly, Table \ref{tab:efftot5} summarizes the relevant results for energies
where the Geant4 parametrized model is applicable.

The experimental data sample encompasses measurements with different
uncertainties; nevertheless, it is worthwhile to note that consistent outcome of
the $\chi^2$ test comparing experimental and calculated cross sections is
observed over the whole range of data precision:
%for instance, the ``pass'' and ``fail'' resulting from tests that involve experimental
%uncertainties in the first quartile of precision (relative uncertainties smaller
%than 2\%) and in the last one (larger than 5\%) are statistically equivalent.
for instance, no statistically significant dissimilarity is present in the ``pass'' and
``fail'' results of tests that involve experimental uncertainties in the first
quartile of precision (relative uncertainties smaller than 2\%) and in the last
one (experimental uncertainties larger than 5\%).
This result suggests that the ability to appraise the capabilities of the
computational models subject to evaluation is not significantly affected by the
variable precision of the experimental sample, and that the presence of 
lower precision measurements is not expected to introduce a 
substantial bias in the categorical data analysis.

% Table generated by Excel2LaTeX from sheet 'pass_range2'
\begin{table}[htbp]
  \centering
  \caption{Outcome of the $\chi^2$ test over total cross sections above 100 eV}
    \begin{tabular}{lccc}
\toprule
    \textbf{Model} & \multicolumn{1}{c}{\textbf{Pass}} & \multicolumn{1}{c}{\textbf{Fail}} & \multicolumn{1}{c}{\textbf{Efficiency}} \\
\midrule
    EPDL97 & 49    & 15    & 0.77 $\pm$ 0.05 \\
    EPICS2017 & 48    & 16    & 0.75 $\pm$ 0.05 \\
\midrule
    Biggs & 48    & 16    & 0.75 $\pm$ 0.05 \\
\midrule
    Pen2011 & 48    & 16    & 0.75 $\pm$ 0.05 \\
    Pen2014 & 27    & 37    & 0.42 $\pm$ 0.06 \\
\midrule
    SabCar & 44    & 20    & 0.69 $\pm$ 0.06 \\
    SabWCar & 44    & 20    & 0.69 $\pm$ 0.06 \\
    SabPCar & 27    & 37    & 0.42 $\pm$ 0.06 \\
    SabWPCar & 27    & 37    & 0.42 $\pm$ 0.06 \\
\midrule
    SabWil & 43    & 21    & 0.67 $\pm$ 0.06 \\
    SabWWil & 42    & 22    & 0.66 $\pm$ 0.06 \\
    SabPWil & 29    & 35    & 0.45 $\pm$ 0.06 \\
    SabWPWil & 29    & 35    & 0.45 $\pm$ 0.06 \\
\midrule
    SabDef & 45    & 19    & 0.70 $\pm$ 0.06 \\
    SabWDef & 44    & 20    & 0.69 $\pm$ 0.06 \\
    SabPDef & 28    & 36    & 0.44 $\pm$ 0.06 \\
    SabWPDef & 28    & 36    & 0.44 $\pm$ 0.06 \\
\midrule
    SabECar & 44    & 20    & 0.69 $\pm$ 0.06 \\
    SabEWCar & 44    & 20    & 0.69 $\pm$ 0.06 \\
    SabEPCar & 27    & 37    & 0.42 $\pm$ 0.06 \\
    SabEWPCar & 27    & 37    & 0.42 $\pm$ 0.06 \\
\midrule
    SabEWil & 43    & 21    & 0.67 $\pm$ 0.06 \\
    SabEWWil & 43    & 21    & 0.67 $\pm$ 0.06 \\
    SabEPWil & 29    & 35    & 0.45 $\pm$ 0.06 \\
    SabEWPWil & 28    & 36    & 0.44 $\pm$ 0.06 \\
\midrule
    SabEDef & 45    & 19    & 0.70 $\pm$ 0.06 \\
    SabEWDef & 44    & 20    & 0.69 $\pm$ 0.06 \\
    SabEPDef & 28    & 36    & 0.44 $\pm$ 0.06 \\
    SabEWPDef & 28    & 36    & 0.44 $\pm$ 0.06 \\
\bottomrule
    \end{tabular}%
  \label{tab:efftot100}%
\end{table}%

% Table generated by Excel2LaTeX from sheet 'pass_range9'
\begin{table}[htbp]
  \centering
  \caption{Outcome of the $\chi^2$ test over parameterized total cross sections  above approximately 5 keV}
    \begin{tabular}{lccc}
 \toprule
   \textbf{Model} & \multicolumn{1}{c}{\textbf{Pass}} & \multicolumn{1}{c}{\textbf{Fail}} & \multicolumn{1}{c}{\textbf{Efficiency}} \\
\midrule
    EPDL97  & 28    & 3     & 0.90 $\pm$ 0.06 \\
%    EPICS2017 & 26    & 5     & 0.84 \\
%\midrule
    Biggs & 28    & 3     & 0.90 $\pm$ 0.06 \\
    Param6 & 19    & 12    & 0.61 $\pm$ 0.08 \\
%\midrule
%    Pen2011 & 28    & 3     & 0.90 \\
%    Pen2014 & 23    & 8     & 0.74 \\
%\midrule
%    SabCar & 27    & 4     & 0.87 \\
%    SabWCar & 27    & 4     & 0.87 \\
%    SabPCar & 23    & 8     & 0.74 \\
%    SabWPCar & 23    & 8     & 0.74 \\
%\midrule
%    SabWil & 27    & 4     & 0.87 \\
%    SabWWil & 26    & 5     & 0.84 \\
%    SabPWil & 23    & 8     & 0.74 \\
%    SabWPWil & 23    & 8     & 0.74 \\
%\midrule
%    SabDef & 28    & 3     & 0.90 \\
%    SabWDef & 27    & 4     & 0.87 \\
%    SabPDef & 23    & 8     & 0.74 \\
%    SabWPDef & 23    & 8     & 0.74 \\
%\midrule
%    SabECar & 27    & 4     & 0.87 \\
%    SabEWCar & 27    & 4     & 0.87 \\
%    SabEPCar & 23    & 8     & 0.74 \\
%    SabEWPCar & 23    & 8     & 0.74 \\
%\midrule
%    SabEWil & 27    & 4     & 0.87 \\
%    SabEWWil & 28    & 3     & 0.90 \\
%    SabEPWil & 23    & 8     & 0.74 \\
%    SabEWPWil & 23    & 8     & 0.74 \\
%\midrule
%    SabEDef & 28    & 3     & 0.90 \\
%    SabEWDef & 27    & 4     & 0.87 \\
%    SabEPDef & 23    & 8     & 0.74 \\
%    SabEWPDef & 23    & 8     & 0.74 \\
\bottomrule
    \end{tabular}%
  \label{tab:efftot5}%
\end{table}%

Qualitatively, above 100 eV one can observe higher efficiencies for EPDL97, EPICS2017, the
original Biggs parameterization, Penelope 2011 and Sabbatucci-Salvat
calculations not involving Pratt's correction.
Substantial differences are visible in the compatibility with experiment of some
categories of computational methods: between the original Biggs parameterization
and the Geant4 6-parameter one, between Penelope 2011 and 2014 cross sections, and regarding
the role of Pratt's correction in Sabbatucci-Salvat calculations.

\begin{figure}[!hbp]
    \centering
    \begin{subfigure}[b]{0.48\columnwidth}
        \includegraphics[width=\textwidth]{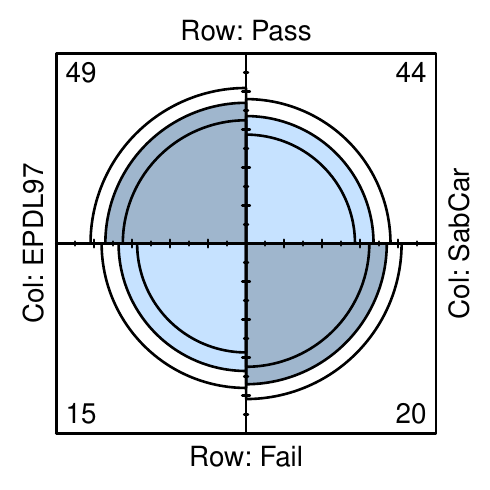}
        \caption{Carlson binding energies}
        \label{fig_SabCar}
    \end{subfigure}%
     \begin{subfigure}[b]{0.48\columnwidth}
        \includegraphics[width=\textwidth]{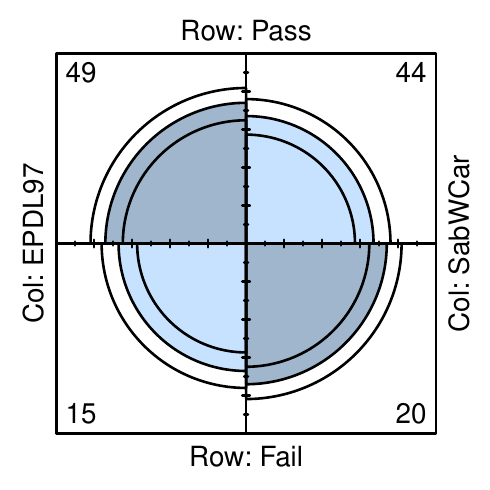}
       \caption{Accounting for level width}
        \label{fig_SabWCar}
    \end{subfigure}
    %add desired spacing between images, e. g. ~, \quad, \qquad, \hfill etc. 
    %(or a blank line to force the subfigure onto a new line)
    \begin{subfigure}[b]{0.48\columnwidth}
        \includegraphics[width=\textwidth]{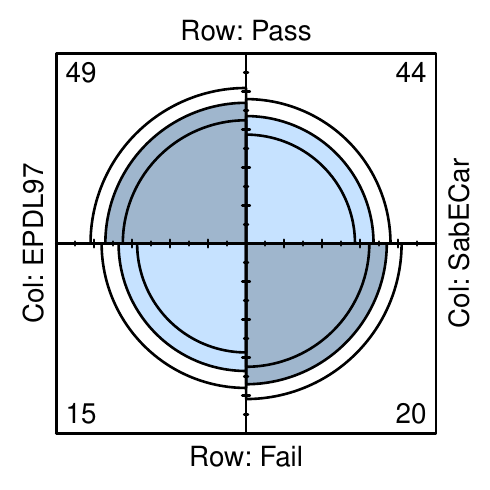}
        \caption{Including excitation to bound levels}
        \label{fig_SabECar}
    \end{subfigure}%
     \begin{subfigure}[b]{0.48\columnwidth}
        \includegraphics[width=\textwidth]{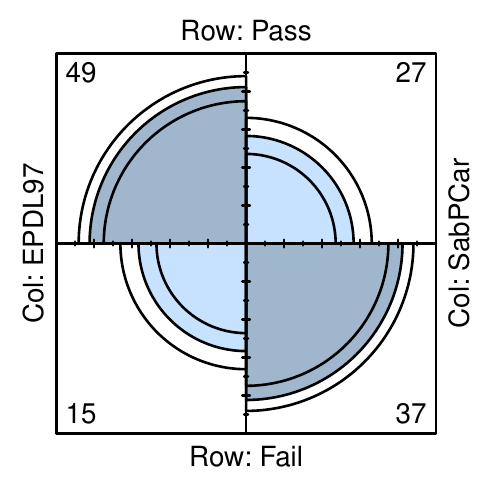}
       \caption{Including Pratt's correction}
        \label{fig_SabPCar}
    \end{subfigure}
\caption{Visual representation of the 2 by 2 table summarizing the compatibility
with experiment of total photoelectric cross sections based on EPDL97 and on
different options of Sabbatucci-Salvat calculations, above 100 eV.}
\label{fig_fourfold_sab}
\end{figure}

\begin{figure}[!htbp]
    \centering
    \begin{subfigure}[b]{0.48\columnwidth}
        \includegraphics[width=\textwidth]{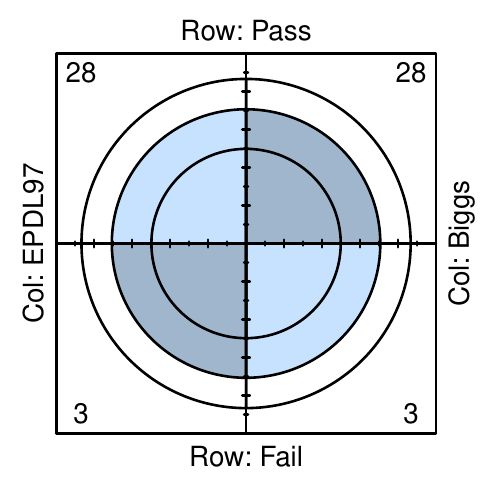}
        \caption{EPDL97 - Biggs}
        \label{fig_biggs}
    \end{subfigure}%
     \begin{subfigure}[b]{0.48\columnwidth}
        \includegraphics[width=\textwidth]{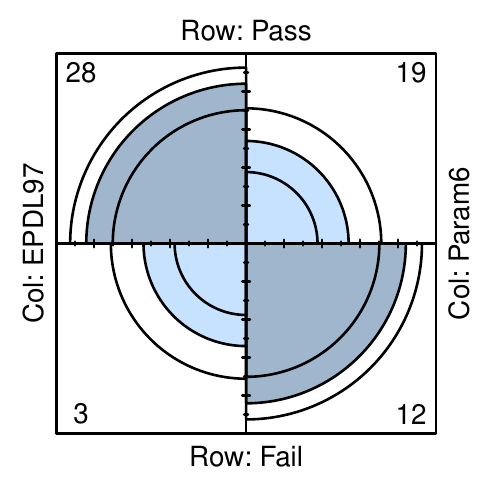}
       \caption{EPDL97 - Param6}
        \label{fig_param6}
    \end{subfigure}
    %add desired spacing between images, e. g. ~, \quad, \qquad, \hfill etc. 
    %(or a blank line to force the subfigure onto a new line)
\caption{Visual representation of the 2 by 2 tables summarizing the compatibility
with experiment of total photoelectric cross sections calculated by EPDL97 and
by parameterized models, above approximately 5~keV.} 
\label{fig_fourfold_param}
\end{figure}

\begin{figure}[!htbp]
\centerline{\includegraphics[angle=0,width=0.48\columnwidth]{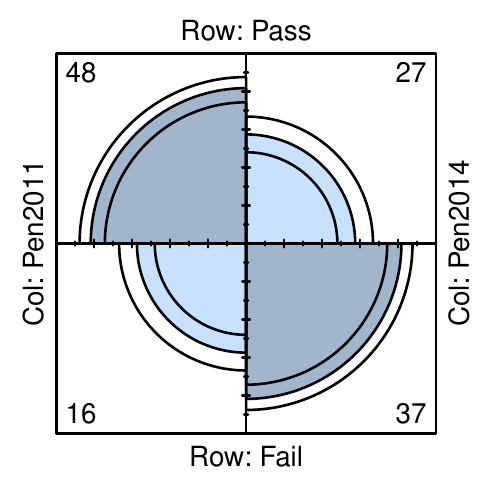}}
\caption{Visual representation of the 2 by 2 table summarizing the compatibility
with experiment of Penelope 2011 and 2014 total photoelectric cross sections.}
\label{fig_fourfold_pen}
\end{figure}

Figs. \ref{fig_fourfold_sab}-\ref{fig_fourfold_pen} highlight some of the
observed differences; they focus on the effect of options of cross section
calculations based on the DHFS method (Fig. \ref{fig_fourfold_sab}), on the performance of the
original Biggs-Lighthill parameterization and of the new one included in Geant4
10.5 (Fig. \ref{fig_fourfold_param}), and on comparing the compatibility with
experiment of Penelope 2011 and 2014 (Fig. \ref{fig_fourfold_pen}).
The so-called \textit{fourfold} plots provide a graphical representation of the contingency tables that
summarize the results of the $\chi^2$ test comparing experimental and calculated
cross sections: the data are standardized so that both table margins are equal,
while preserving the odds ratio; the area of each quarter circle is proportional
to the standardized cell frequency in the corresponding table.
An association (odds ratio different from~1) between the row and column
variables of the table is indicated by the tendency of diagonally opposite cells in one
direction to differ in size from those in the other direction.
The innermost and outermost ring show the limits of a 99\% confidence interval for the odds ratio; overlapping
rings in adjacent quadrants indicate consistency with the null hypothesis of
independence, i.e. the hypothesis that the different ``pass'' and ``fail''
observed in the two categories could arise from chance only.
Details about fourfold plots can be found in
textbooks such as \cite{friendly2000, friendly2015}.

\begin{figure*}[!htbp]
    \centering
    \begin{subfigure}[b]{0.48\columnwidth}
        \includegraphics[width=\textwidth]{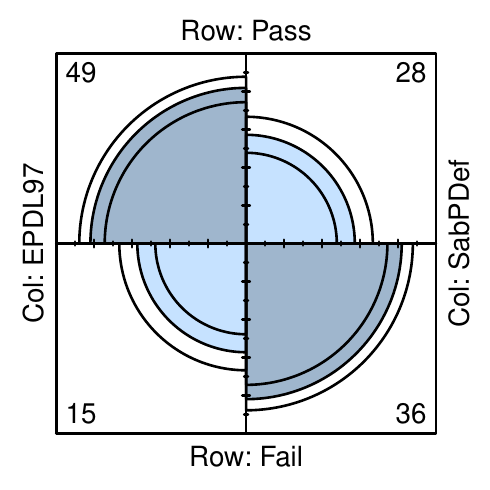}
        \caption{E $>$ 100 eV}
        \label{fig_pratt100}
    \end{subfigure}%
     \begin{subfigure}[b]{0.48\columnwidth}
        \includegraphics[width=\textwidth]{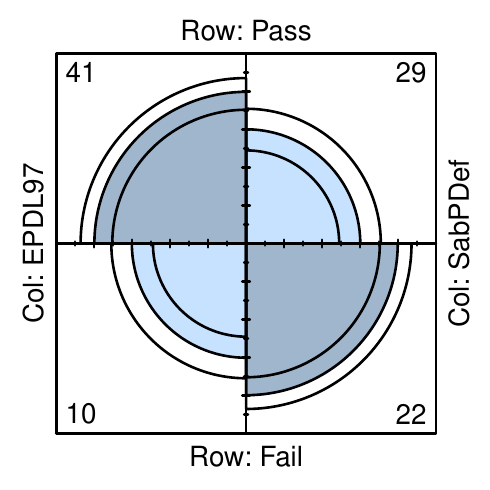}
       \caption{E $>$ 250 eV}
        \label{fig_pratt250}
    \end{subfigure}%
   \centering
    \begin{subfigure}[b]{0.48\columnwidth}
        \includegraphics[width=\textwidth]{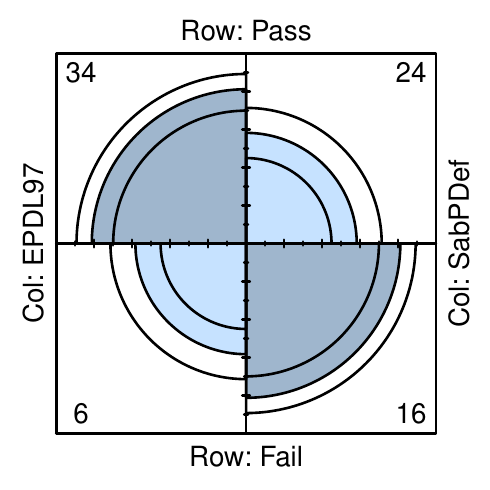}
        \caption{E $>$ 1 keV}
        \label{fig_pratt1}
    \end{subfigure}%
\caption{Visual representation of 2 by 2 tables summarizing the compatibility
with experiment of total photoelectric cross sections based on EPDL97 and
on  Sabbatucci-Salvat calculations including Pratt's correction,
concerning different energies.} 
\label{fig_fourfold_pratt}
\end{figure*}

Figs. \ref{fig_SabCar}, \ref{fig_SabWCar} and \ref{fig_SabECar} represent
contingency tables that compare the compatibility with experiment of EPDL97 with
that of three variants of Sabbatucci-Salvat calculations using Carlson’s binding
energies: plain calculations (identified as SabCar), calculations accounting for
finite level widths (SabWCar) and calculations accounting for excitation to
bound states (SabECar).
These  plots visually convey the message that these three calculation
configurations result in similar compatibility with experiment w.r.t. EPDL97.
%this qualitative observation is confirmed by the tests on the corresponding
%contingency tables in Table . 
Fig. \ref{fig_SabPCar} compares the compatibility
with experiment of EPDL97 and Sabbatucci-Salvat calculations using the same
binding energies and applying the normalization correction à la Pratt (SabPCar);
it qualitatively shows that this
correction appears responsible for more pronounced discrepancy in compatibility
with experiment w.r.t. EPDL97.
%This qualitative observation is confirmed by the rejection of the null hypothesis of equivalent compatibility with experiment in the tests of the corresponding contingency table, documented in Table \ref{tab:contTotEPDL}.
A similar effect is observed in  Fig. \ref{fig_fourfold_pen}, which concerns
Penelope 2014, using cross sections derived from Sabbatucci-Salvat calculations, and Penelope 2011,
using EPDL97.

These indications are objectively quantified by applying the tests listed in
Section \ref{sec:anal} to the corresponding contingency tables.
Since EPDL97 exhibits the largest efficiency among the computational methods
subject to evaluation (as in \cite{tns_photoel}), the statistical analysis compares 
its performance with respect to experiment with that of other calculations.

The results of the statistical analysis of contingency tables are summarized in Table
\ref{tab:contTotEPDL} for energies above 100 eV.
The hypothesis of equivalent compatibility with experiment with respect to
EPDL97 is rejected by all tests 
% with 0.01 significance 
for Penelope 2014 cross
sections and for Sabbatucci-Salvat calculations involving Pratt's normalization
correction; it is not rejected for the other computational methods.
Presumably, the difference in compatibility with experiment between Penelope
2011 and Penelope 2014 is related to the application of Pratt's correction in
the latter.
It is worthwhile to note that the null hypothesis is rejected with 0.001 significance level, 
which exceeds the significance set a priori in the context of this validation study.

The rejection occurs with 0.001 significance not only over the whole data
sample, but even over data samples involving experimental precisions above
the median, i.e. encompassing measurements with $> 3\%$ relative uncertainty: that is, the difference
in compatibility with experiment with respect to EPDL97 is observed also in the
presence of larger experimental uncertainties.
This observation confirms that the presence of lower precision measurements in the
data sample does not hinder the capability to identify
statistically significant dissimilarities in the calculated cross sections.

% Table generated by Excel2LaTeX from sheet 'contingency_range2_EPDL'
\setlength{\tabcolsep}{5pt}
\begin{table}[!htbp]
  \centering
  \caption{P-values of tests  comparing the compatibility with experiment of total cross sections 
of EPDL97 and of other computational models, for energies above 100 eV}
    \begin{tabular}{lccccc}
\toprule
    \textbf{Models} & \textbf{Fisher} & \textbf{$\chi^2$} & \textbf{Boschloo} & \textbf{Z-pooled} & \textbf{CSM} \\
\midrule
    EPICS2017 & 1.000 & 0.837 & 1.000 & 0.905 & 0.997 \\
\midrule
    Biggs & 1.000 & 0.837 & 1.000 & 0.905 & 0.997 \\
\midrule
   Pen2011 & 1.000 & 0.837 & 1.000 & 0.905 & 0.997 \\
    Pen2014 & $< 0.001$ & $< 0.001$ & $< 0.001$ & $< 0.001$ & $< 0.001$ \\
\midrule
    SabCar & 0.428 & 0.321 & 0.338 & 0.529 & 0.405 \\
    SabWCar & 0.428 & 0.321 & 0.338 & 0.529 & 0.405 \\
    SabPCar & $< 0.001$ & $< 0.001$ & $< 0.001$ & $< 0.001$ & $< 0.001$ \\
    SabWPCar & $< 0.001$ & $< 0.001$ & $< 0.001$ & $< 0.001$ & $< 0.001$ \\
\midrule
    SabWil & 0.326 & 0.238 & 0.250 & 0.250 & 0.308 \\
    SabWWil & 0.242 & 0.172 & 0.185 & 0.243 & 0.226 \\
    SabPWil & 0.001 & $< 0.001$ & $< 0.001$ & $< 0.001$ & $< 0.001$ \\
    SabWPWil & 0.001 & $< 0.001$ & $< 0.001$ & $< 0.001$ & $< 0.001$ \\
\midrule
    SabDef & 0.549 & 0.423 & 0.448 & 0.531 & 0.601 \\
    SabWDef & 0.428 & 0.321 & 0.338 & 0.529 & 0.405 \\
    SabPDef & $< 0.001$ & $< 0.001$ & $< 0.001$ & $< 0.001$ & $< 0.001$ \\
    SabWPDef & $< 0.001$ & $< 0.001$ & $< 0.001$ & $< 0.001$ & $< 0.001$ \\
\midrule
    SabECar & 0.428 & 0.321 & 0.338 & 0.529 & 0.405 \\
    SabEWCar & 0.428 & 0.321 & 0.338 & 0.529 & 0.405 \\
    SabEPCar & $< 0.001$ & $< 0.001$ & $< 0.001$ & $< 0.001$ & $< 0.001$ \\
    SabEWPCar & $< 0.001$ & $< 0.001$ & $< 0.001$ & $< 0.001$ & $< 0.001$ \\
\midrule
    SabEWil & 0.326 & 0.238 & 0.250 & 0.250 & 0.308 \\
    SabEWWil & 0.326 & 0.238 & 0.250 & 0.250 & 0.308 \\
    SabEPWil & 0.001 & $< 0.001$ & $< 0.001$ & $< 0.001$ & $< 0.001$ \\
    SabEWPWil & $< 0.001$ & $< 0.001$ & $< 0.001$ & $< 0.001$ & $< 0.001$ \\
\midrule
    SabEDef & 0.549 & 0.423 & 0.448 & 0.531 & 0.601 \\
    SabEWDef & 0.428 & 0.321 & 0.338 & 0.529 & 0.405 \\
    SabEPDef & $< 0.001$ & $< 0.001$ & $< 0.001$ & $< 0.001$ & $< 0.001$ \\
    SabEWPDef & $< 0.001$ & $< 0.001$ & $< 0.001$ & $< 0.001$ & $< 0.001$ \\
\bottomrule
    \end{tabular}%
  \label{tab:contTotEPDL}%
\end{table}%
\setlength{\tabcolsep}{6pt}

\begin{figure}[!htbp]
    \centering
    \begin{subfigure}[b]{0.48\columnwidth}
        \includegraphics[width=\textwidth]{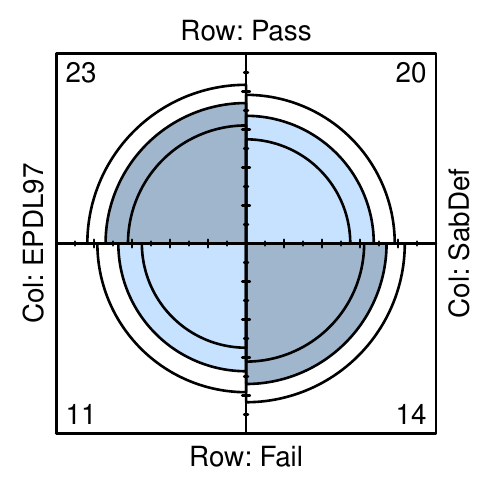}
        \caption{Without normalization correction}
        \label{fig_nopratt100}
    \end{subfigure}%
     \begin{subfigure}[b]{0.48\columnwidth}
        \includegraphics[width=\textwidth]{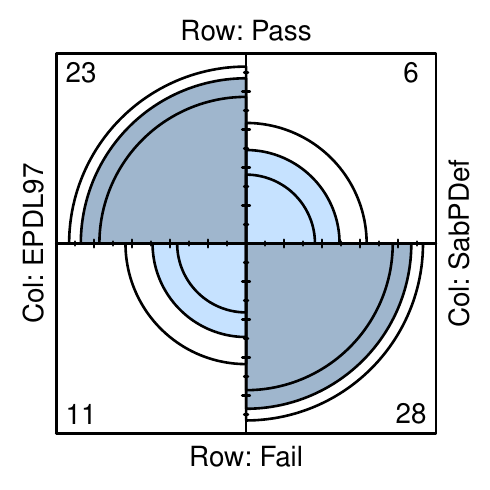}
       \caption{With normalization correction}
        \label{fig_pratt100}
    \end{subfigure}
    %add desired spacing between images, e. g. ~, \quad, \qquad, \hfill etc. 
    %(or a blank line to force the subfigure onto a new line)
\caption{Visual representation of the 2 by 2 tables summarizing the compatibility
with experiment of total photoelectric cross sections calculated by EPDL97 and
by Sabbatucci-Salvat, including or not including Pratt's rnormalization correction. The results 
concern data in the energy range  between 100~eV and 1~keV.} 
\label{fig_fourfold_prattlow}
\end{figure}

Fig. \ref{fig_fourfold_pratt} suggests that the effect of the normalization
correction is more pronounced when  lower energy data are involved.
The plots concern data above 100 eV, 250 eV and 1 keV; they illustrate the
compatibility with experiment of total photoelectric cross sections based on
EPDL97 and on Sabbatucci-Salvat calculations including Pratt's correction, 
using the default atomic binding energies of the PHOTACS code.
This effect is clearly visible in Fig. \ref{fig_fourfold_prattlow}, which concerns 
data between 100~eV and 1~keV.

For energies between 100~eV and 1~keV, the hypothesis of equivalent compatibility
with experiment with respect to EPDL97 is rejected for all Sabbatucci-Salvat
calculation options including the normalization correction, and for Penelope
2014; it is not rejected for the other computational methods.
The statistical analysis of the contingency tables associated with Fig. \ref{fig_fourfold_pratt}, reported in Table
\ref{tab:pratt}, shows that the null hypothesis of equivalent compatibility with
experiment is rejected with 0.01 significance for the sample involving cross
sections down to 100 eV, while it is not rejected for the data samples at higher
energies.
Nevertheless, the null hypothesis is rejected with 0.05 significance over the
higher energy data samples.
This result suggests that the inadequacy of calculations involving the
normalization correction to some extent persists even above 1~keV,
which is conventionally considered a conservative lower limit for the
reliability for photon transport in general-purpose Monte Carlo codes.

 % Table generated by Excel2LaTeX from sheet 'contingency_range4_EPDL'
\begin{table}[!htbp]
  \centering
  \caption{P-values of tests comparing the compatibility with experiment of total cross sections 
based on EPDL97 and on Sabbatucci-Salvat calculations including Pratt's correction, 
over different energies}
%of EPDL97 and of other computational models, for energies above 1 keV}}
    \begin{tabular}{lccccc}
\toprule
    \textbf{Energy} & \textbf{Fisher} & \textbf{$\chi^2$} & \textbf{Boschloo} & \textbf{Z-pooled} & \textbf{CSM} \\
\midrule
     E $>$ 100 eV & $< 0.001$ & $< 0.001$ & $< 0.001$ & $< 0.001$ & $< 0.001$ \\
     E $>$ 250 eV & 0.018 & 0.010 & 0.013 & 0.013 & 0.011 \\
     E $>$ 1 keV & 0.023 & 0.012 & 0.014 & 0.013 & 0.015 \\
\bottomrule
    \end{tabular}%
  \label{tab:pratt}%
\end{table}%

\begin{table}[!htbp]
  \centering
    \centering
  \caption{P-values of tests comparing the compatibility with experiment of total cross sections 
of EPDL97 and of parameterized models, for energies above approximately  5 keV}
    \begin{tabular}{lcccc}
  \toprule
    \textbf{Models} & \textbf{Fisher}  & \textbf{Boschloo} & \textbf{Z-pooled} & \textbf{CSM} \\
\midrule    
%    EPICS2017 & 0.707 & 0.526 & 0.530 & 0.995 \\
% \midrule
    Biggs & 1 & 1 & 1 & 1 \\
    Param6 & 0.0159 & 0.0080 & 0.0080 & 0.0097 \\
%\midrule
%    Pen2011 & 1.000 & 1.000 & 1.000 & 1.000 \\
%    Pen2014 & 0.182 & 0.113 & 0.116 & 0.138 \\
%\midrule
%    SabCar & 1.000 & 1.000 & 0.789 & 1.000 \\
%    SabWCar & 1.000 & 1.000 & 0.789 & 1.000 \\
%    SabPCar & 0.182 & 0.113 & 0.116 & 0.138 \\
%    SabWPCar & 0.182 & 0.113 & 0.116 & 0.138 \\
% \midrule
%   SabWil & 1.000 & 1.000 & 0.789 & 1.000 \\
%    SabWWil & 0.707 & 0.526 & 0.530 & 0.995 \\
%    SabPWil & 0.182 & 0.113 & 0.116 & 0.138 \\
%    SabWPWil & 0.182 & 0.113 & 0.116 & 0.138 \\
% \midrule
%   SabDef & 1.000 & 1.000 & 1.000 & 1.000 \\
%    SabWDef & 1.000 & 1.000 & 0.789 & 1.000 \\
%    SabPDef & 0.182 & 0.113 & 0.116 & 0.138 \\
%    SabWPDef & 0.182 & 0.113 & 0.116 & 0.138 \\
%\midrule
%    SabECar & 1.000 & 1.000 & 0.789 & 1.000 \\
%    SabEWCar & 1.000 & 1.000 & 0.789 & 1.000 \\
%    SabEPCar & 0.182 & 0.113 & 0.116 & 0.138 \\
%    SabEWPCar & 0.182 & 0.113 & 0.116 & 0.138 \\
%\midrule
%    SabEWil & 1.000 & 1.000 & 0.789 & 1.000 \\
%    SabEWWil & 1.000 & 1.000 & 1.000 & 1.000 \\
%    SabEPWil & 0.182 & 0.113 & 0.116 & 0.138 \\
%    SabEWPWil & 0.182 & 0.113 & 0.116 & 0.138 \\
%\midrule
%    SabEDef & 1.000 & 1.000 & 1.000 & 1.000 \\
%    SabEWDef & 1.000 & 1.000 & 0.789 & 1.000 \\
%    SabEPDef & 0.182 & 0.113 & 0.116 & 0.138 \\
%    SabEWPDef & 0.182 & 0.113 & 0.116 & 0.138 \\
\bottomrule
    \end{tabular}%
 \label{tab:contTotEPDL5keV}%
\end{table}%

The analysis of cross section parameterizations is summarized in Table
\ref{tab:contTotEPDL5keV}, which reports the results of the tests comparing the
compatibility with experiment of EPDL97 total cross sections and of
parameterized models above approximately 5~keV.
The hypothesis of equivalent compatibility with experiment of EPDL97 and of
Biggs-Lighthill original parameterization is not rejected by any of the tests
applied to the corresponding contingency table.
Regarding the parameterization included in Geant4 10.5, the null hypothesis is
rejected with 0.01 significance by Boschloo \cite{boschloo}, Z-pooled
\cite{suissa} and CSM-approximated Barnard \cite{barnard} test, while it is not
rejected by the Fisher test, which is known to be more conservative over
$2\times 2$ tables \cite{agresti1992}.
The rejection occurs even over the data samples involving experimental precisions above
the median, i.e. encompassing measurements with $> 3\%$ relative uncertainty: that is, the difference
in compatibility with experiment with respect to EPDL97 does not appear to be
influenced by lower precision measurements.

% -----------------------------------------------------------------------------------------
\subsection{Partial Cross Sections}

\begin{figure}[!htbp]
\centerline{\includegraphics[angle=0,width=7.9cm]{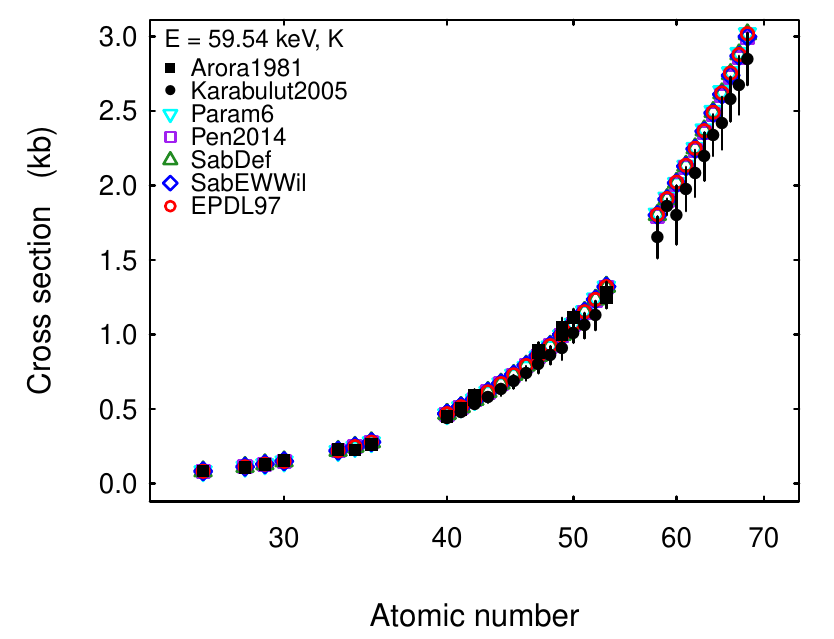}}
\caption{K shell photoionization cross section at 59.54~keV as a function of the atomic number Z.}
\label{fig_K59}
\end{figure}

\begin{figure}[!htbp]
\centerline{\includegraphics[angle=0,width=7.9cm]{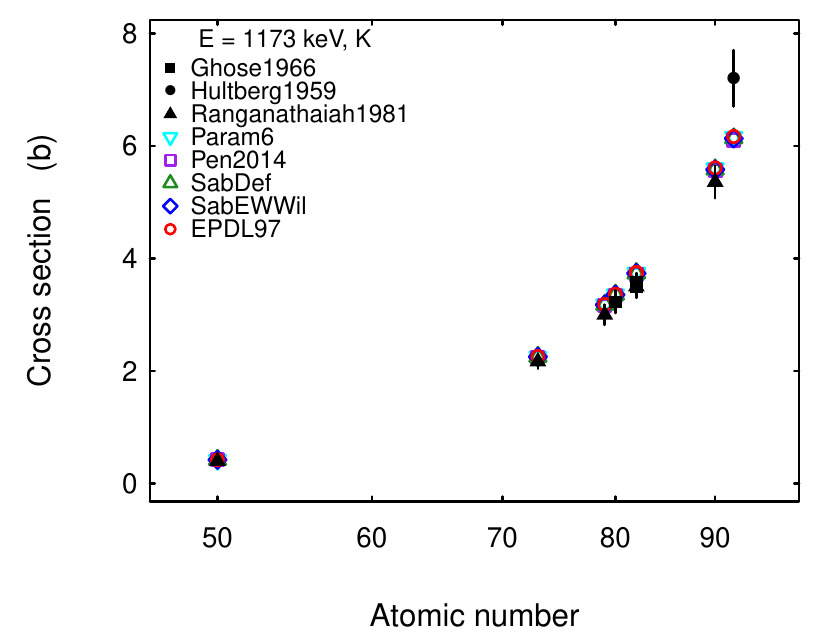}}
\caption{K shell photoionization cross section at 1.173 MeV as a function of the atomic number Z.}
\label{fig_K1173}
\end{figure}

Figs. \ref{fig_K59}-\ref{fig_K1173} represent  qualitative examples
of how EPDL97 and some more recent computational methods reproduce experimental
K shell photoelectric cross sections.


The outcome of the $\chi^2$ test comparing calculated and experimental K shell
cross sections  is summarized in Table \ref{tab:effK100} for data samples above 100~eV
and in Table \ref{tab:effK5} for data above approximately 5~keV
All the models involved in the test qualitatively exhibit similar capabilities to 
reproduce experimental measurements.

% Table generated by Excel2LaTeX from sheet 'pass_range2'
\begin{table}[!htbp]
  \centering
  \caption{Outcome of the $\chi^2$ test over K shell cross sections above 100~eV}
    \begin{tabular}{lccc}
\toprule
    \textbf{Model} & \textbf{Pass} & \textbf{Fail} & \textbf{Efficiency} \\
    EPDL97 & 23    & 3     & 0.88 $\pm$ 0.06 \\
    EPICS2017 & 24    & 2     & 0.92 $\pm$ 0.06 \\
\midrule
    Pen2011 & 24    & 2     & 0.92 $\pm$ 0.06 \\
    Pen2014 & 23    & 3     & 0.88 $\pm$ 0.06 \\
 \midrule
   SabCar & 23    & 3     & 0.88 $\pm$ 0.06 \\
    SabWCar & 24    & 2     & 0.92 $\pm$ 0.06 \\
    SabPCar & 23    & 3     & 0.88 $\pm$ 0.06 \\
    SabWPCar & 24    & 2     & 0.92 $\pm$ 0.06 \\
\midrule
    SabWil & 23    & 3     & 0.88 $\pm$ 0.06 \\
    SabWWil & 24    & 2     & 0.92 $\pm$ 0.06 \\
    SabPWil & 23    & 3     & 0.88 $\pm$ 0.06 \\
    SabWPWil & 24    & 2     & 0.92 $\pm$ 0.06 \\
\midrule
    SabDef & 23    & 3     & 0.88 $\pm$ 0.06 \\
    SabWDef & 23    & 3     & 0.88 $\pm$ 0.06 \\
    SabPDef & 23    & 3     & 0.88 $\pm$ 0.06 \\
    SabWPDef & 23    & 3     & 0.88 $\pm$ 0.06 \\
 \midrule
   SabECar & 23    & 3     & 0.88 $\pm$ 0.06 \\
    SabEWCar & 24    & 2     & 0.92 $\pm$ 0.06 \\
    SabEPCar & 23    & 3     & 0.88 $\pm$ 0.06 \\
    SabEWPCar & 24    & 2     & 0.92 $\pm$ 0.06 \\
 \midrule
   SabEWil & 23    & 3     & 0.88 $\pm$ 0.06 \\
    SabEWWil & 25    & 1     & 0.96 $\pm$ 0.06 \\
    SabEPWil & 23    & 3     & 0.88 $\pm$ 0.06 \\
    SabEWPWil & 25    & 1     & 0.96 $\pm$ 0.06 \\
\midrule
    SabEDef & 23    & 3     & 0.88 $\pm$ 0.06 \\
    SabEWDef & 23    & 3     & 0.88 $\pm$ 0.06 \\
    SabEPDef & 23    & 3     & 0.88 $\pm$ 0.06 \\
    SabEWPDef & 23    & 3     & 0.88 $\pm$ 0.06 \\
\bottomrule
    \end{tabular}%
  \label{tab:effK100}%
\end{table}%

% Table generated by Excel2LaTeX from sheet 'pass_range9'
\begin{table}[htbp]
  \centering
  \caption{Outcome of the $\chi^2$ test over relevant  K shell cross sections  above approximately 5 keV}
    \begin{tabular}{lccc}
 \toprule
   \textbf{Model} & \multicolumn{1}{c}{\textbf{Pass}} & \multicolumn{1}{c}{\textbf{Fail}} & \multicolumn{1}{c}{\textbf{Efficiency}} \\
\midrule
    EPDL97  	& 22    & 3     	& 0.88 $\pm$ 0.07 \\
    EPICS2017 	& 23    & 2     	& 0.92 $\pm$ 0.06 \\
    Param6 		& 21    & 4		& 0.84 $\pm$ 0.07 \\
\bottomrule
    \end{tabular}%
  \label{tab:effK5}%
\end{table}%

The analysis of contingency tables does not identify any significant differences
between the compatibility with experiment of K shell cross sections calculated
by EPDL97 and by other computational methods.
The results for data above 100~eV are summarized in Table
\ref{tab:contKEPDL}: the hypothesis of equivalent compatibility with experiment
with respect to EPDL97 is not rejected for any of the alternative computational
methods.
It is not rejected either in the tests concerning the domain of applicability of
the Geant4 10.5 parameterized model above approximately 5~keV.
However, it is worthwhile to note that the smaller size of the K shell data
sample reduces the power of the test with respect to the analysis of total cross
sections: for instance, the power of the Boschloo test to detect the observed
difference between total cross sections calculated by EPDL97 and with Pratt's
correction is 0.93, while it drops to 0.48 for observing the same effect 
in the K shell data sample.

% Table generated by Excel2LaTeX from sheet 'contingency_range2_EPDL'
\begin{table}[!htbp]
  \centering
  \caption{P-values of tests  comparing the compatibility with experiment of K shell cross sections 
of EPDL97 and of other computational models, for energies above 100 eV}
    \begin{tabular}{lcccc}
 \toprule
    \textbf{Models} & \textbf{Fisher}  & \textbf{Boschloo} & \textbf{Z-pooled} & \textbf{CSM} \\
\midrule   
    EPICS2017 & 1.000 & 1.000 & 0.749 & 1.000\\    
% \midrule   
 %  Param6 & 0.703 & 0.629 & 0.714 & 0.998 \\
\midrule   
    Pen2011 & 1.000 & 1.000 & 0.749 & 1.000 \\
    Pen2014 & 1.000 & 1.000 & 1.000 & 1.000 \\
\midrule   
    SabCar & 1.000 & 1.000 & 1.000 & 1.000 \\
    SabWCar & 1.000 & 1.000 & 0.749 & 1.000 \\
    SabPCar & 1.000 & 1.000 & 1.000 & 1.000 \\
    SabWPCar & 1.000 & 1.000 & 0.749 & 1.000 \\
\midrule   
    SabWil & 1.000 & 1.000 & 1.000 & 1.000 \\
    SabWWil & 1.000 & 1.000 & 0.749 & 1.000 \\
    SabPWil & 1.000 & 1.000 & 1.000 & 1.000 \\
    SabWPWil & 1.000 & 1.000 & 0.749 & 1.000 \\
 \midrule   
    SabDef & 1.000 & 1.000 & 1.000 & 1.000 \\
    SabWDef & 1.000 & 1.000 & 1.000 & 1.000 \\
    SabPDef & 1.000 & 1.000 & 1.000 & 1.000 \\
    SabWPDef & 1.000 & 1.000 & 1.000 & 1.000 \\
\midrule   
    SabECar & 1.000 & 1.000 & 1.000 & 1.000 \\
    SabEWCar & 1.000 & 1.000 & 0.749 & 1.000 \\
    SabEPCar & 1.000 & 1.000 & 1.000 & 1.000 \\
    SabEWPCar & 1.000 & 1.000 & 0.749 & 1.000 \\
\midrule   
    SabEWil & 1.000 & 1.000 & 1.000 & 1.000 \\
    SabEWWil & 0.610 & 0.488 & 0.359 & 1.000 \\
    SabEPWil & 1.000 & 1.000 & 1.000 & 1.000 \\
    SabEWPWil & 0.610 & 0.488 & 0.359 & 1.000 \\
\midrule   
    SabEDef & 1.000 & 1.000 & 1.000 & 1.000 \\
    SabEWDef & 1.000 & 1.000 & 1.000 & 1.000 \\
    SabEPDef & 1.000 & 1.000 & 1.000 & 1.000 \\
    SabEWPDef & 1.000 & 1.000 & 1.000 & 1.000 \\
\bottomrule
    \end{tabular}%
  \label{tab:contKEPDL}%
\end{table}%

The scarcity of experimental data hinders a meaningful validation analysis for L
subshell cross sections and for outer shells.
Regarding L subshells, all computational methods exhibit the same compatibility
with experiment with the exception of the Geant4 10.5 parameterization \`a la
Biggs; nevertheless, the observed differences are not statistically significant.
It was already remarked in \cite{tns_photoel} that the experimental data sample
for outer shells is inadequate to perform validation tests.

% -----------------------------------------------------------------------------------------
\section{Conclusion}

Statistical tests against a large sample of photoelectric cross section
measurements allow a rigorous and objective characterization of the available
computational methods and identification of the state of the art in this
domain.

The validation tests documented in this paper
%involving a large experimental data sample collected from the literature, 
demonstrate that EPDL97 cross section
tabulations, based on Scofield's 1973 calculations, still
represent the state of the art for total and K shell photoelectric cross
sections: there is no evidence that more recent computational methods surpass EPDL97
compatibility with experimental data.
The new version of EPDL, released in ENDF/B-VIII.0, exhibits statistically equivalent 
behaviour with respect to total and K shell photoelectric cross sections.
EPDL97 is used by several Monte Carlo particle transport codes; the results of
this paper show that there is no need to move to more recent data libraries for
the simulation of the photoelectric effect.

The detailed formulation of the theory documented in \cite{sabbatucci_2016}
allows the evaluation of various physics options.
The tests fail to reject the hypothesis of 
equivalent compatibility with experiment with respect to EPDL97 when accounting for finite level widths and excitation to bound states.
No significantly different validation results are observed by adjusting the
theoretically calculated cross sections according to either Carlson's or Williams'
compilations of atomic binding energies.

The normalization applied to calculations adopting the DHFS approach according to \cite{sabbatucci_2016}, based on the
multi-configuration code by Desclaux, appears to deteriorate the accuracy of
total cross sections: goodness of fit tests show that the hypothesis of
compatibility with experimental data is rejected in a larger number of test
cases, resulting in statistically significant differences in reproducing experimental 
measurements with respect to EPDL97.
The effect of deterioration is not visible in the tests concerning the K shell;
nevertheless, the power of these tests is lower due to the smaller experimental
data sample involved.
Caution should be exercised in using this correction, properly identifying the 
limits of its applicability.

The results of the validation analysis suggest that the application of the
normalization correction could be responsible for the apparently degraded
compatibility with experiment of Penelope 2014 photoelectric cross sections
with respect to the previous 2011 version.

The replacement of EPDL97 interpolation in Geant4 10.5 \textit{low energy}
electromagnetic package with an empirical parameterization \`a la
Biggs-Ligthhill above approximately 5~keV significantly degrades the
compatibility of total cross sections with experiment.
The original Biggs-Lighthill parameterization, used in Geant4 \textit{standard}
electromagnetic package, is not affected by this drawback in the same energy
range.

The conceptual framework of hypothesis testing \cite{conover} does not allow discerning whether
the failure to reject the null hypothesis 
% of equivalent compatibility with experiment 
is due to the hypothesis being ``true'' or to insufficient evidence
from the data to reject it.
% inadequacy of the data
Therefore, more extensive experimental measurements are needed to quantify the
capabilities of partial cross section calculations and of other features, such
as accounting for excitation to bound levels and finite level widths, which could not be discriminated
with respect to EPDL97 on the basis of currently available experimental data.
High-precision measurements in the proximity of absorption edges,
with reliable estimates of their uncertainties, would also be helpful to better
characterize the capabilities of the various computational methods.

The results of the validation tests documented in this paper provide guidance to
the developers and users of Monte Carlo particle transport codes to optimize the 
choice of photoelectric cross sections in the simulation of experimental scenarios.

% -----------------------------------------------------------------------------------------

% ------------------------------------------------------------------------------
\section*{Acknowledgment}

The authors express their gratitude to Francesc Salvat, who kindly provided 
tabulations resulting from
\cite{sabbatucci_2016} and the \textit{PHOTACS-PP} code implementing further 
computational options on top of them.
%The authors express their gratitude to CERN for support to the research
%described in this paper.
The authors thank Anita Hollier for proofreading the manuscript and Min Cheol
Han for his interest in the early stage of the validation project.
The CERN Library has provided helpful assistance and essential reference 
material for the validation tests.
The author Tullio Basaglia wishes to specify that his contribution to this paper concerns
the cross section data.

% ------------------------------------------------------------------------


\begin{thebibliography}{199}

\bibitem{tns_photoel}
M. C. Han et al., 
``Validation of Cross Sections for Monte Carlo Simulation of the Photoelectric Effect'',
\emph{IEEE Trans. Nucl. Sci.}, vol. 63, no. 2, pp. 1117-1146, 2016.

% 

\bibitem{epdl97}	
D. Cullen et al., 
``EPDL97, the Evaluated Photon Data Library'', 
Lawrence Livermore Natl. Lab. Report UCRL-50400, Vol. 6, Rev. 5, 1997.

\bibitem{scofield_1973}
J. H. Scofield, 
``Theoretical photoionization cross sections from 1 to 1500 keV'',
%Report
 UCRL-51326, Lawrence Livermore Lab., CA, USA, 1973.

\bibitem{andreobook}
P. Andreo, D. T. Burns, A. E. Nahum, J. Seuntjens, F. H. Attix, 
``Fundamentals of Ionizing Radiation Dosimetry'', Wiley, Chapter 3.3, 2017.

\bibitem{sabbatucci_2016}
L. Sabbatucci and F. Salvat,
``Theory and calculation of the atomic photoeffect'',
\textit{Radiat. Phys. Chem.}, vol. 121, pp. 122-140, 2016.

\bibitem{penelope2014}
F. Salvat, J. M. Fernandez-Varea and J. Sempau,
``Penelope-2014 - A code system for Monte Carlo simulation of electron and
photon transport'', 
Proc. Workhop NEA/NSC/DOC(2015)3, 2015.

\bibitem{epdl2017}
D. E. Cullen,``A Survey of Photon Cross Section Data for use in EPICS2017, rev. 1'',
IAEA-NDS-0225, Vienna, Austria, 2017.

\bibitem{biggs1}
F. Biggs and R. Lighthill,
``Analytical Approximations For X-ray Cross-sections'',
Sandia National Laboratories Report SC-RR-66-452, Albuquerque, 1967.

\bibitem{biggs2}
F. Biggs and R. Lighthill,
``Analytical Approximations For X-ray Cross-sections II'',
Sandia National Laboratories Report SC-RR-710507, Albuquerque, 1971.

\bibitem{biggs3}
F. Biggs and R. Lighthill,
``Analytical Approximations for X-RayCrossSections III''
Sandia National Laboratories Report SAND87-0070, Albuquerque, 1988.

\bibitem{geantv}
G. Amadio et al., 
``GeantV alpha release'',
\textit{J. Phys. Conf. Series}, vol. 1085, p. 032037, 2018.

% Basic Geant4 references
\bibitem{g4nim} 
S.~Agostinelli et al., 
``Geant4 - a simulation toolkit''
\textit{Nucl. Instrum. Meth. A}, vol. 506, no. 3, pp. 250-303, 2003.

\bibitem{g4tns}
J.~Allison et al., 
``Geant4 Developments and Applications'' 
\textit{IEEE Trans. Nucl. Sci.}, vol. 53, no. 1, pp. 270-278, 2006.

\bibitem{g4nim2}
J. Allison et al.,
``Recent developments in Geant4'',
\textit{Nucl. Instrum. Meth. A}, vol. 835, pp. 186-225, 2016.

% Used by Sabbatucci
\bibitem{pratt_1960b}
R. H. Pratt, 
``Photoeffect from the L shell'',
\textit{Phys. Rev.}, vol. 119, no. 5, pp. 1619–1626, 1960.

\bibitem{pratt_1973}
R. H. Pratt, A. Ron and H. K. Tseng, 
``Atomic photoelectric effect above 10 keV'',
\textit{Rev. Mod. Phys.},  vol.  45, no. 2, pp. 273-325, 1973.

\bibitem{pratt_1973_err}
 R. H. Pratt, A. Ron and H. K. Tseng, 
``Erratum: Atomic photoelectric effect above 10 keV'',
\textit{Rev. Mod. Phys.},  vol.  45, no. 4, pp. 663–664, 1973.

\bibitem{desclaux_1975}
J. P. Desclaux,
``A multiconfiguration relativistic Dirac-Fock program'',
\textit{Comp. Phys. Comm.}, vol. 9, no. 1, pp. 31–45, 1975.

\bibitem{desclaux_1977}
J. P. Desclaux,
``Erratum notice'',
 textit{Comp. Phys. Comm.}, vol. 13, no. 1, p. 71, 1977.

\bibitem{salvat_2018}
 F. Salvat and X. Llovet,
``Monte Carlo simulation and fundamental quantities'',
\textit{IOP Conf. Ser.: Mater. Sci. Eng.}, vol. 304, p. 012014, 2018.

\bibitem{pratt_1960a}
R. H. Pratt, 
``Atomic photoelectric effect at high energies'',
\textit{Phys. Rev.}, vol. 117, pp. 1017–1028, 1960.

\bibitem{carlson_1975}
T. A. Carlson, 
``Photoelectron and Auger Spectroscopy'', 
Plenum Press, New York, 1975.

\bibitem{penelope2011}
F. Salvat, J.M. Fernandez-Varea, and J. Sempau,
``Penelope-2011 - A code system for Monte Carlo simulation of electron and photon transport'', 
Proc. Workhop NEA/NSC/DOC(2011)5, 2011.

% EPICS 2017
\bibitem{endfb8}
D. A. Brown et al.,
``ENDF/B-VIII.0: The 8$^{th}$ Major Release of the Nuclear Reaction Data Library with CIELO-project Cross Sections, New Standards and Thermal Scattering Data'',
\textit{Nucl. Data Sheets}, vol. 148, pp. 1-142, 2018.

\bibitem{epics2017}
D. E. Cullen, EPICS2017, Online.  
Available: https://www-nds.iaea.org/epics/.

\bibitem{eadl}
S. T. Perkins et al., 
``Tables and Graphs of Atomic Subshell and
Relaxation Data Derived from the LLNL Evaluated Atomic Data Library (EADL) Z=1-100'', 
UCRL-50400 Vol. 30, Lawrence Livermore Natl. Lab., 1991.

\bibitem{eadl2017}
D. E. Cullen,
``A Survey of Atomic Binding Energies for use in EPICS2017'',
IAEA-NDS-0224, Vienna, Austria, 2017.

\bibitem{tns_epics2017}
M. C. Han et al.,
``First Assessment of ENDF/B-VIII and EPICS Atomic Data Libraries'',
\textit{IEEE Trans. Nucl. Sci.}, vol. 63, no. 8, pp. 2268-2278, 2018.

\bibitem{endfb8_errata}
ENDF/B-VIII.0 Errata, Online.  
Available: \url{https://www.nndc.bnl.gov/endf/b8.0/errata.html}.

\bibitem{g4physrefmanual10.5}
Geant4 Collaboration, 
``Geant4 Physics Reference Manual, Geant4 version 10.5'', [Online]. 
%Available: http://geant4.web.cern.ch/\-geant4/\-User\-Documentation/\-Users\-Guides/\-Physics\-Reference\-Manual/\-fo/\-Physics\-Reference\-Manual.pdf.
Available:  http://cern.ch/\-geant4-userdoc/\-UsersGuides/\-Physics\-Reference\-Manual/\-fo/\-Physics\-Reference\-Manual.pdf

\bibitem{epics2014}
D. Cullen et al., 
``EPICS2014: Electron Photon Interaction Cross Section (Version 2014)'', 
IAEA-NDS-218, rev.1, 2015.

%% ---- Verification and Validation
%
\bibitem{ieee_vv}
%IEEE Computer Society,
``IEEE Standard for System, Software, and Hardware Verification and Validation'',
IEEE Std 1012-2016, pp.1-260, 2017.

\bibitem{williams_2011}
G. P. Williams,
``Electron binding energies of the elements'',
 in \textit{CRC Handbook of Chemistry and Physics}, 91st edition, ed.  W. M. Haynes and D. R. Lide, CRC, Boca Raton, pp. 221–226, 2011.


\bibitem{refactoring}
M. Fowler, 
``Refactoring: improving the design of existing code'',  Boston: Addison-Wesley, 1999.

%\bibitem{neyman_1928}
%J. Neyman and E. S.  Pearson.
%''On the use and interpretation of certain test criteria for purposes of statistical inference: part I'',
%\textit {Biometrika}, vol. 20A, pp. 175–240, 1928.

%  ---- End of photoel2 bibliography

%\bibitem{scofield_1973}
%J. H. Scofield
%\textit{Theoretical photoionization cross sections from 1 to 1500 keV},
%. No. UCRL--51326. California Univ., Livermore. Lawrence Livermore Lab., 1973.
%Lawrence Livermore Laboratory Rep. UCRL-51326, 1973.
%

%\bibitem{cesareo1992} 
%R. Cesareo, A. L. Hanson, G. E. Gigante, L. J. Pedraza, and S. Q. G. Mahtaboally,
%``Interaction of keV photons with matter and new applications'',
%\textit{Phys. Rep.}, vol. 213, no. 3, pp. 117-178, 1992.

% Reviews


% Basic Geant4 references:
%\bibitem{g4nim} 
%S.~Agostinelli et al., 
%``Geant4 - a simulation toolkit''
%\textit{Nucl. Instrum. Meth. A}, vol. 506, no. 3, pp. 250-303, 2003.
%
%\bibitem{g4tns}
%J.~Allison et al., 
%``Geant4 Developments and Applications'' 
%\textit{IEEE Trans. Nucl. Sci.}, vol. 53, no. 1, pp. 270-278, 2006.

%\bibitem{tns_relax_prob}
%M. G. Pia, P. Saracco, and M. Sudhakar,
%``Validation of K and L Shell Radiative Transition Probability Calculations'',
%\textit{IEEE Trans. Nucl. Sci.}, vol. 56, no. 6, pp. 3650-3661,  2009.
%
%\bibitem{tns_binding}
%M. G. Pia et al.,
%``Evaluation of atomic electron binding energies for Monte Carlo particle transport'',
%\textit{IEEE Trans. Nucl. Sci.}, vol. 58, no. 6, pp. 3246-3268, 2011.
%
%
%% Biggs
%
%
%
%\bibitem{epdl78}
%E. F. Plechaty, D. E. Cullen, and R. J. Howerton, 
%``Tables and Graphs of Photon-Interaction Cross Sections From 0.1 keV to 100 MeV Derived from the LLL Evaluated-Nuclear-Data Library'', 
%UCRL-50400, Vol. 6, Rev. 2, Livermore National Laboratory, December 7, 1978).
%
%\bibitem{grishin_1994}
%V. M. Grishin, A. P. Kostin, S. K. Kotelnikov and D. G. Streblechenko,
%Parametrization of the photoabsorption cross section at low energies'',
%\textit{Bull. Lebedev Inst.}, no. 6, pp. 1-6, 1994. 
%
%\bibitem{ENDFBVI}
%C. L. Dunford,
%``Evaluated Nuclear Data File, ENDF/B-VI'',
%in \textit{Nuclear Data for Science and Technology}, pp. 788-792, Springer, Berlin, 1992.
%i%n \textit{Nuclear Data for Science and Technology, Research Reports in Physics}, pp. 788-792, 1992.
%%Nuclear Data for Science and Technology
%
%
%
%
%\bibitem{ENDFBVII}
%M. B. Chadwick et al.,
%``ENDF/B-VII.1 Nuclear Data for Science and Technology: Cross Sections,
%Covariances, Fission Product Yields and Decay Data'',
%\textit{Nucl. Data Sheets}, vol. 112, no. 12, pp. 2887-3152, 2011.
%
%\bibitem{epdl81}
%E. F. Plechaty, D. E. Cullen, and R. J. Howerton, 
%``Tables and Graphs of Photon-Interaction Cross Sections From 0.1 keV to 100 MeV Derived from the LLL Evaluated-Nuclear-Data Library'', 
%UCRL-50400, Vol. 6, Rev. 3, Livermore National Laboratory, 1981).
%
%\bibitem{epics2017}
%D. Cullen et al., 
%``A Survey of Photon Cross Section Data for use in EPICS2017'', 
%IAEA-NDS-225, 2017.
%
%
%
%% MC codes
%
%% EGS
%
%\bibitem{egs4}
%W. R. Nelson, H. Hirayama, and D. W. O. Rogers, 
%``The EGS4 Code System'',
%SLAC-265 Report, Stanford, CA, 1985. 
%
%\bibitem{sakamoto}
%Y. Sakamoto,
%``Photon cross section data PHOTX for PEGS4 code'',
% in \textit{Proc.Third EGS4 User’s Meeting in Japan}, KEK Proceedings 93-15, pp. 77-82, Japan, 1993.
%
%\bibitem{egsnrc}
%I. Kawrakow, E. Mainegra-Hing, D.W.O. Rogers, F. Tessier and B. R. B. Walters, 
%``The EGSnrc Code System: Monte Carlo Simulation of Electron and Photon Transport'',
%NRCC PIRS-701 Report, 6th printing,  National Research Council of Canada, Ottawa, Canada, 2013. 
%
%
%% FLUKA
%\bibitem{fluka1}
%G. Battistoni et al.,
%``The FLUKA code: description and benchmarking'',
%\textit{AIP Conf. Proc.}, vol. 896, pp. 31-49, 2007.
%%A.~Fass{\`o} et al., ``The physics models of FLUKA: status and recent
%%developments'', in {\sl Proc.\ Computing in High Energy and Nuclear
%%Physics 2003 Conference (CHEP 2003)}, La Jolla, CA, USA, paper MOMT05.
%
%\bibitem{fluka2}
%A.~Ferrari et al., 
%``Fluka: a multi-particle transport code'', 
%Report CERN-2005-010, INFN/TC-05/11, SLAC-R-773, Geneva, Oct. 2005.
%
%% ITS
%\bibitem{its5}
%B. C. Franke, R. P. Kensek and T. W. Laub,
%``ITS5 theory manual'', rev. 1.2,
%Sandia Natl. Lab. Report SAND2004-4782, Albuquerque, 2004.
%
%
%% MCNP
%
%\bibitem{epdl89}
%D. E. Cullen et al.,
%``Tables and Graphs of Photon Interaction Cross Sections from 10 eV to 100 GeV
%Derived from the LLNL Evaluated Photon Data Library (EPDL)'',
%Lawrence Livermore National Laboratory Report UCRL-50400, Vol. 6, Rev. 4, 1989.
%
%\bibitem{mcnp6}
%T. Goorley et al., 
%``Initial MCNP6 Release Overview - MCNP6 version 1.0'', 
%LA-UR-13-22934 Report, Los Alamos National Laboratory, 2013.
%
%\bibitem{mcnp6_2}
%T. Goorley et al., 
%``MCNP Version 6.2 Release Notes'', 
%LA-UR-18-20808 Report, Los Alamos National Laboratory, 2018.
%
%% Penelope
%
%%\bibitem{penelope}
%%J.~Baro, J.~Sempau, J. M.~Fern\'andez-Varea, and F.~Salvat,
%%``PENELOPE, an algorithm for Monte Carlo simulation of the
%%penetration and energy loss of electrons and positrons in
%%matter'', 
%%\emph{Nucl. Instrum. Meth. B}, vol. 100, no. 1, pp. 31-46, 1995.
%
%\bibitem{sempau_1997}
%J. Sempau, E. Acosta, J. Baro, J. M. Fernandez-Varea, and F. Salvat,
%``An algorithm for Monte Carlo simulation of coupled electron-photon transport'',
%\emph{Nucl. Instrum. Meth. B}, vol. 132, pp. 377-390, 1997.
%
%%\bibitem{penelope2001}
%%F. Salvat, J.M. Fernandez-Varea, E. Acosta, and J. Sempau,
%%``Penelope - A code system for Monte Carlo simulation of electron and
%%photon transport'', 
%%Proc. Workshop NEA, 2001.
%
%\bibitem{penelope2008}
%F. Salvat, J. M. Fernandez-Varea, and J. Sempau,
%``Penelope - A code system for Monte Carlo simulation of electron and
%photon transport'', 
%Proc. Workshop NEA 6416, 2008.
%
%
%
%
%\bibitem{g4physmanual_1003}
%Geant4 Collaboration, 
%``Geant4 Physics Reference Manual version 10.03'', [Online]. 
%Available: http://geant4.web.cern.ch/\-geant4/\-UserDocumentation/\-UsersGuides/\-PhysicsReferenceManual/\-BackupVersions/\-V10.3/\-fo/\-PhysicsReferenceManual.pdf.
%
%\bibitem{g4physmanual_1004}
%Geant4 Collaboration, 
%``Geant4 Physics Reference Manual'', [Online]. 
%Available: http://geant4.web.cern.ch/\-geant4/\-User\-Documentation/\-Users\-Guides/\-Physics\-Reference\-Manual/\-fo/\-Physics\-Reference\-Manual.pdf.
%
%
%
%\bibitem{g4pai}
%J. Apostolakis, S. Giani, L. Urbanb, M. Maire, A. V. Bagulya, and V. M. Grichine, 
%``An implementation of ionisation energy loss in very thin absorbers for the GEANT4 simulation package'',
%\emph{Nucl. Instrum. Meth. A}, vol. 453, no. 3, pp. 597-605, 2000.
%
%
%\bibitem{lowe_chep}
%S. Chauvie, G. Depaola, V. Ivanchenko, F. Longo, P. Nieminen and M. G. Pia,
%``Geant4 Low Energy Electromagnetic Physics'',
%in \textit{Proc. Computing in High Energy and Nuclear Physics}, 
%Beijing, China, pp. 337-340, 2001.
%
%\bibitem{lowe_nss}
%S. Chauvie et al., ``Geant4 Low Energy Electromagnetic Physics'',
%in \textit{2004 IEEE Nucl. Sci. Symp. Conf. Rec.}, pp. 1881-1885, 2004.
%
%\bibitem{lowe_e} 
%J. Apostolakis, S. Giani, M. Maire, P. Nieminen, M.G. Pia, L. Urban,
%``Geant4 low energy electromagnetic models for electrons and photons'',
%\textit{INFN/AE-99/18}, Frascati, 1999. 
%
%
%

%
%% ---- Experimental data
%
%% ---- Start total experimental
%

%
%\bibitem{iso12207}
%ISO/IEC, ``International Standard, Information Technology Software Life
%Cycle Process, IS0 12207'', 2008-2011.
%
%
%
%% ---- Statistics
%
%\bibitem{gof1}
%G. A. P. Cirrone et al., 
%``A Goodness-of-Fit Statistical Toolkit'', 
%\emph{IEEE Trans. Nucl. Sci.}, vol. 51, no. 5, pp. 2056-2063, 2004.
%
%\bibitem{gof2}
%B. Mascialino, A. Pfeiffer, M. G. Pia, A. Ribon, and P. Viarengo, 
%``New developments of the Goodness-of-Fit Statistical Toolkit'', 
%\emph{IEEE Trans. Nucl. Sci.}, vol. 53, no. 6, pp. 3834-3841,  2006.
%

\bibitem{bock}
R. K. Bock and W. Krischer,
``The Data Analysis BriefBook '',
Ed. Springer, Berlin, 1998. 

\bibitem{fisher}
R. A. Fisher,
``On the interpretation of  $\chi^2$ from contingency tables, and the calculation of P'',
\textit{J. Royal Stat. Soc.}, vol. 85, no. 1, pp. 87-94, 1922. 

\bibitem{boschloo}
R. D. Boschloo, 
``Raised Conditional Level of Significance for the 2$\times$2-table when Testing the Equality of Two Probabilities'', 
\textit{Stat. Neerlandica}, vol. 24, pp. 1-35, 1970.

\bibitem{suissa}
S. Suissa,  and  J. J. Shuster,
``Exact Unconditional Sample Sizes for the 2$\times$2 Binomial Trial'',
\textit{J. Royal Stati. Soc., Ser. A}, vol. 148, pp. 317-327, 1985.

%\bibitem{santner}
%T. J. Santner and M. K. Snell,
%``Small-Sample Confidence Intervals for p$_1$ - p$_2$ and p$_{1}/$p$_2$ in 2$\times$2 Contingency Tables
%Journal of the American Statistical Association
%Vol. 75, No. 370 (Jun., 1980), pp. 386-394 
%
\bibitem{barnard}
 G. A. Barnard,
``Significance tests for 2$\times$2 tables'',
\textit{Biometrika}, vol. 34, pp. 123-138, 1947.

\bibitem{barnard_csm}
E. B. Wilson,
``Barnard's CSM Test of Significance'',
\textit{Proc. Natl. Acad. Sci.  United States of America}, vol. 38, no. 10, pp. 899-905, 1952.

\bibitem{pearson}
K. Pearson,
``On the $\chi^2$ test of Goodness of Fit'',
\textit{Biometrika}, vol. 14, no. 1-2, pp. 186-191, 1922.

\bibitem{lyonsbook}
L. Lyons,
``Statistics for Nuclear and Particle Physicists'',
Cambridge University Press, 1989.

\bibitem{R}
R Core Team,
``R: A language and environment for statistical computing'', 
R Foundation for Statistical Computing, Vienna, Austria, 2019. 
[Online]. Available: http://www.R-project.org/.

\bibitem{friendly2000}
M. Friendly, 
``Visualizing Categorical Data'', Section 3.4,  SAS Institute Inc., Cary, NC, USA, 2000, 

\bibitem{friendly2015}
M. Friendly and D. Meyer, 
``Discrete Data Analysis with R: Visualization and Modeling Techniques for Categorical and Count Data'', Section 4.4, 
CRC Press, Boca Raton, FL, USA, 2015.

\bibitem{agresti1992}
A. Agresti, 
``A Survey of Exact Inference for Contingency Tables'', 
\textit{Stat. Sci.}, vol. 7, pp. 131-153, 1992.

\bibitem{conover}
W.  J. Conover, 
``Practical Nonparametric Statistics'', 
3$^{rd}$ ed., Chapter 2, 
Wiley, 1999.
%
%\bibitem{mato_1997}
%S. Mato and M. Andres, 
%``Simplifying the calculation of the P-value for Barnard’s test and its derivatives'',
%\textit{Stat. Comp.}, vol. 7, pp. 137-143, 1997.
%
%\bibitem{agresti}
%A. Agresti, 
%``A Survey of Exact Inference for Contingency Tables'', 
%\textit{Stat. Sci.}, vol. 7, pp. 131-153, 1992.
%
%\bibitem{andres_1994}
%A. Martin Andres and  A. Silva Mato,   
%``Choosing the optimal unconditioned test for comparing two independent proportions'',
%\textit{Comp. Stat. Data Anal.},  vol. 17, pp. 555-574, 1994.
%
%\bibitem{andres_2004}
%A. Martin Andres, A. Silva Mato, J. M . Tapia Garcia, and M. J. Sanchez Quevedo,  
%``Comparing the asymptotic power of exact tests in 2 x 2 tables'',
%\textit{Comp. Stat. Data Anal.},  vol.  47, pp. 745-756, 2004.
%
%%\bibitem{eadl}
%%S. T. Perkins et al., 
%%``Tables and Graphs of Atomic Subshell and
%%Relaxation Data Derived from the LLNL Evaluated Atomic Data Library (EADL)'', 
%%Z=1-100, UCRL-50400 Vol. 30, 1997.
%%
%%\bibitem{eadl}
%%S. T. Perkins et al., 
%%``Tables and Graphs of Atomic Subshell and
%%Relaxation Data Derived from the LLNL Evaluated Atomic Data Library (EADL)'', 
%%Z=1-100, UCRL-50400 Vol. 30, 1997.
%
%%\bibitem{pratt_1973}
%%R. H. Pratt, A Ron, and H. K. Tseng, 
%%``Atomic photoelectric effect above 10 keV",
%%\textit{Mod. Phys.},  vol.  45, pp. 273-325, 1973.

\end{thebibliography}
\end{document}